\documentclass[fleqn,10pt]{wlscirep}
\usepackage[utf8]{inputenc}
\usepackage[T1]{fontenc}
\usepackage{authblk} 
\usepackage{tabularx}
\usepackage{multirow} 
\title{ CFM-GP: Unified Conditional Flow Matching to Learn Gene Perturbation Across Cell Types}

\author[1,\dag]{Abrar Rahman Abir}
\author[2,\dag]{Sajib Acharjee Dip}
\author[2,*]{Liqing Zhang}

\affil[1]{Bangladesh University of Engineering and Technology, Bangladesh}
\affil[2]{Virginia Tech, USA}

\affil[*]{Corresponding authors: \texttt{abrarrahmanabir156@gmail.com}, \texttt{lqzhang@cs.vt.edu}}
\affil[$\dagger$]{These authors contributed equally to this work.}

\begin{abstract}
Understanding gene perturbation effects across diverse cellular contexts is a central challenge in functional genomics, with significant implications for therapeutic discovery and precision medicine. While single-cell technologies enable high-resolution measurement of transcriptional responses, collecting such data remains expensive and time-intensive, especially when repeated for each cell type. Existing computational methods attempt to predict these responses but typically require separate models per cell type, limiting scalability and generalization.
To address these limitations, we introduce CFM-GP, for cell type–agnostic gene perturbation prediction. Our method learns a continuous, time-dependent transformation between unperturbed and perturbed gene expression distributions, conditioned on cell type. This allows a single model to predict the transcriptional effect of a perturbation across all cell types, eliminating the need for cell type–specific training. Unlike previous approaches that rely on discrete modeling, CFM-GP employs the flow matching objective to model perturbation dynamics in a scalable manner.
We evaluate our approach on five datasets: SARS-CoV-2–infected cells, IFN-$\beta$–stimulated peripheral blood mononuclear cells (PBMC), glioblastoma patients treated with Panobinostat, lupus patients under IFN-$\beta$ stimulation, and Statefate, which examines how cytokines and environment shape progenitor fate. Our model consistently outperforms state-of-the-art baselines, achieving higher R squared scores and Spearman rank correlations, indicating CFM-GP accurately captures the relative ordering of gene expression changes. Moreover, pathway enrichment analysis confirms that our predictions recover key biological pathways activated by perturbations. These findings highlight the robustness and biological fidelity of CFM-GP, establishing its potential as a scalable solution for gene perturbation prediction across cell types.
\end{abstract}
\begin{document}

\flushbottom
\maketitle
%
%
\thispagestyle{empty}


\section*{Introduction}

Gene perturbation refers to the deliberate disruption or modification of a gene's function to observe resulting cellular changes. These interventions provide causal insights into gene regulatory networks by revealing downstream effects on transcription, signaling, and cell fate. Perturbations can be genetic, such as gene knockouts using CRISPR/Cas9 \cite{ja2014genome,wang2022genome}, knockdowns via RNA interference (RNAi) \cite{conte2015rna,fire1998potent}, or overexpression constructs—or chemical, involving small molecules that modulate specific pathways or proteins \cite{szalai2023application}. The transcriptional response elicited by these perturbations serves as a functional fingerprint, reflecting both direct and indirect regulatory consequences and aiding in the identification of essential genes, pathway architecture, and therapeutic targets \cite{szalai2023application,ishikawa2023renge, dip2025moxgate}. When combined with single-cell technologies, perturbation experiments uncover context-specific effects and cellular heterogeneity that bulk profiling may obscure, offering unprecedented resolution in understanding the molecular logic of gene function.

Understanding how gene perturbations reshape transcriptional states is central to functional genomics, drug discovery, and precision medicine. Techniques such as CRISPR knockouts—especially in pooled Perturb‑Seq screens—enable high-throughput exploration of gene function at single‑cell resolution, identifying regulatory networks and therapeutic targets directly from heterogeneous cell populations \cite{dixit2016perturb,park2025perturbomics}. Integrating CRISPR perturbations with scRNA‑seq not only reveals major expression changes but also captures causal, temporal regulation across cell types \cite{ishikawa2023renge}. These technologies have fundamentally transformed our ability to map the transcriptional consequences of genetic or environmental perturbations with cell-type resolution, enabling causal inference of gene function in complex systems. \cite{gavriilidis2024mini}.

The advent of single-cell RNA sequencing (scRNA-seq) has enabled researchers to dissect perturbation responses at unprecedented resolution, capturing heterogeneous and cell-type–specific gene expression changes across complex tissues and conditions. This high granularity offers a powerful lens into the cellular programs affected by genetic or environmental interventions. Large-scale datasets such as scPerturb \cite{green2022scperturb}, Perturb-seq \cite{dixit2016perturb} and CROP-seq \cite{datlinger2017pooled} have provided rich testbeds for modeling these responses, but remain far from covering the full perturbation–cell type space.

However, realizing the full potential of perturbation-based profiling is hampered by significant experimental bottlenecks. Generating perturbation-response data across a broad range of genes, doses, time points, and especially across diverse cell types is prohibitively expensive and time-intensive. Each additional condition requires laborious experimental design, batch handling, and sequencing, making large-scale perturbation screens infeasible in many biological and clinical settings. As a result, the vast combinatorial space of perturbations and cellular contexts remains sparsely sampled, leaving critical gaps in our understanding of gene function across cell types. This motivates the need for scalable computational models that can generalize perturbation effects across unseen conditions and cell identities—serving as powerful surrogates for costly experiments and enabling in silico hypothesis generation.

To overcome data limitations, several computational models have been developed to predict the transcriptional effects of gene perturbations. Early approaches such as scGen \cite{lotfollahi2019scgen}, CVAE \cite{sohn2015learning}, and trVAE \cite{lotfollahi2019conditional} use latent space arithmetic to model expression shifts from control to perturbed states. While effective in some cases, these models assume static perturbation effects and often require retraining for each cell type. scDist \cite{nicol2024robust} offers a statistically robust way to identify affected cell types but cannot generate perturbed profiles or support in silico simulations. Methods like scGen \cite{lotfollahi2019scgen} and CVAE \cite{sohn2015learning} struggle to capture complex, nonlinear, and cell-type–specific responses due to their reliance on linear operations and fixed priors. trVAE improves generalization using MMD regularization, but it lacks explicit cell-type conditioning. scPreGAN \cite{wei2022scpregan} enhances distributional realism through adversarial training, yet its dual-GAN setup is hard to train and does not scale well to new or low-data conditions. CoupleVAE \cite{wu2025couplevae} improves over scGen and CVAE by learning nonlinear bidirectional mappings between control and perturbed states, offering greater flexibility. However, unlike CFM-GP’s unified, cell-type–conditioned framework, it relies on separate encoders and decoders, limiting generalization to unseen cell types. More recent approaches such as CPA \cite{lotfollahi2023predicting} and MultiCPA \cite{inecik2022multicpa} extend this paradigm to compositional settings, allowing for extrapolation across drug dosages and combinations. However, these methods still rely on discrete latent operations, and their generalization across diverse biological contexts—especially across unseen cell types—remains limited.

Orthogonal to latent generative approaches, optimal transport (OT)-based models such as CellOT \cite{bunne2023learning} and sc-OTGM \cite{demir2024sc} have introduced a more principled framework for mapping distributions between control and perturbed populations. These models frame perturbation as a mass-preserving transformation, offering improved distributional fidelity. Yet, most OT-based methods require separate mappings for each cell type and perturbation pair, and they often lack an explicit mechanism to model the temporal progression of gene expression dynamics.

Recent developments in flow-based generative modeling have shown promise for addressing these gaps. Flow matching techniques learn time-dependent vector fields that transform source to target distributions in a continuous, simulation-free manner. While models like CellFlow ~\cite{klein2025cellflow} have applied flow matching to simulate cell phenotypes under perturbation, they do not explicitly incorporate cell type as a conditioning factor and are not optimized for perturbation prediction across diverse biological settings.

To overcome these limitations, we introduce CFM-GP, a unified, conditional flow-matching framework for predicting gene perturbation effects across cell types. CFM-GP models the transformation from unperturbed to perturbed gene expression profiles as a continuous trajectory in expression space, governed by a neural vector field conditioned on cell type. This enables the model to learn cell-type–specific dynamics using a single shared architecture, without requiring retraining for each context.

Our key contributions include:
\begin{itemize}
    \item We develop a novel vector field formulation that learns time-dependent perturbation trajectories between paired control and perturbed single-cell profiles, conditioned on both cell identity and time step. By conditioning on cell type embeddings, CFM-GP generalizes across cellular identities, supporting broad application without the need for cell-type–specific models.
    
    \item We benchmark CFM-GP on five biologically diverse datasets—including SARS-CoV-2 infection \cite{lotfollahi2022mapping}, IFN-\(\beta\) stimulation, glioblastoma drug treatment, systemic lupus erythematosus, and cytokine-driven progenitor fate transitions from the Statefate collection. Our method consistently outperforms baseline models across key evaluation metrics, including \(R^2\), Maximum Mean Discrepancy (MMD) \cite{gretton2012kernel}, and Spearman correlation, capturing both global distributional structure and fine-grained gene expression dynamics.
    
    \item Through gene set enrichment analysis (GSEA) \cite{subramanian2005gene}, we demonstrate that CFM-GP recovers biologically meaningful pathway activity, validating its predictions at a systems level. We also show that CFM-GP successfully transfers learned perturbation dynamics across species boundaries—highlighting its utility in translational and comparative genomics.
\end{itemize}

\section*{Results}
\begin{figure}[htbp]
\centering
\includegraphics[width=\textwidth]{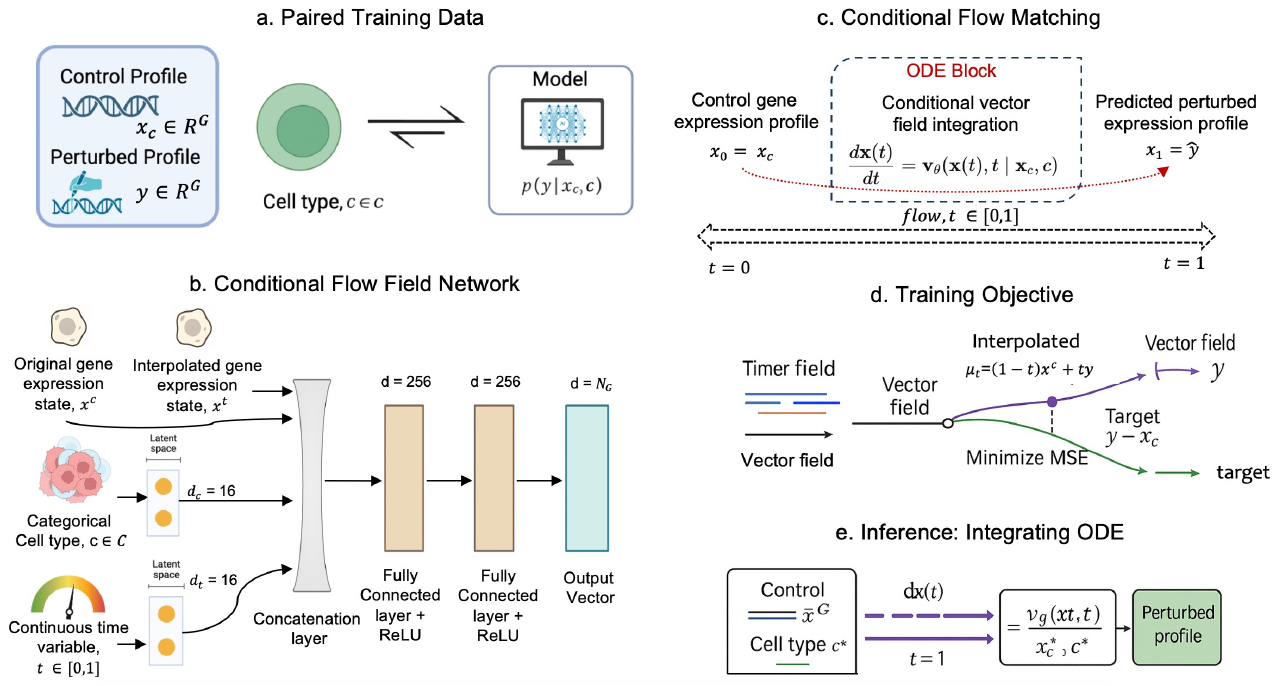}
\caption{
\textbf{Overview of the CFM-GP modeling framework.}
\textbf{(a)} During training, the model receives paired gene expression profiles of individual cells in control and perturbed conditions, along with their associated cell type labels. 
\textbf{(b)} A conditional vector field network \( \mathbf{v}_\theta(\mathbf{x}(t), t \mid \mathbf{x}_c, c) \) is parameterized by a neural network, where each time step receives the interpolated state \( \mathbf{x}(t) \), original control state \( \mathbf{x}_c \), cell type embedding \( c \), and time embedding \( t \).
\textbf{(c)} Conditional Flow Matching drives the learning of continuous trajectories between the control and perturbed states by matching the learned vector field to the ground-truth direction.
\textbf{(d)} The model is trained to minimize the mean squared error between the predicted and true velocity vectors across random interpolations between control and perturbed states.
\textbf{(e)} At inference, the trained vector field is integrated as an ODE from \( t = 0 \) to \( t = 1 \), starting from a new control profile \( \mathbf{x}_c \) and cell type \( c \), to produce a predicted perturbed profile \( \hat{\mathbf{y}} \). 
}
\label{fig:method_pipeline}
\end{figure}

\subsection*{Unified Modeling Across Cell Types by Treating Cell Type as a Condition for Gene Perturbation Prediction}
\begin{figure}[htbp]
    \centering
    \includegraphics[width=0.95\textwidth]{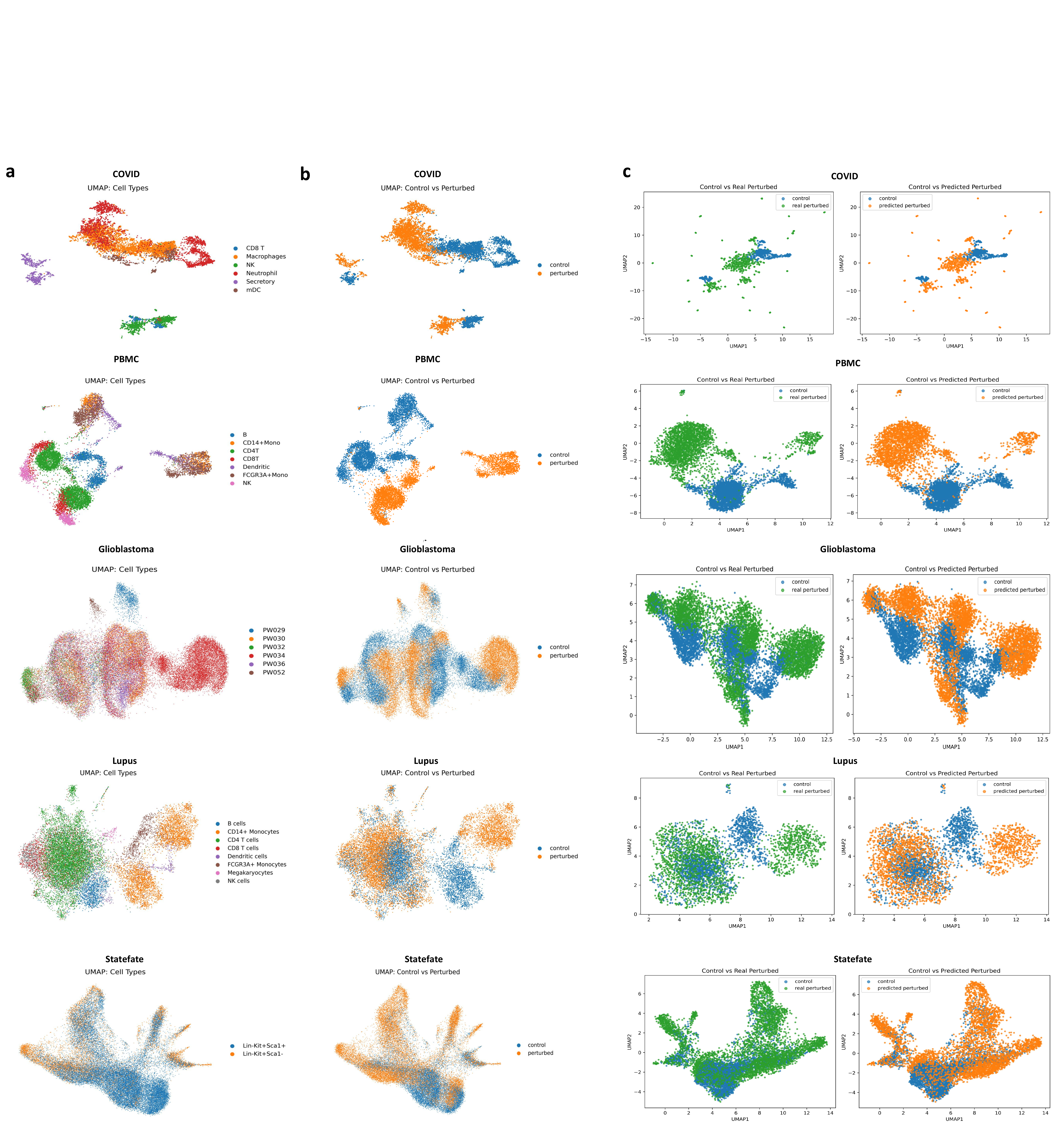}
    \caption{UMAP visualizations illustrating CFM-GP’s ability to preserve cell population structure and accurately reproduce perturbation-induced distribution shifts across multiple datasets. (a) UMAP visualization of ground-truth gene expression profiles across five datasets (COVID, PBMC, Glioblastoma, Lupus, Statefate), showing distinct cell population structures.  
    (b) UMAP visualization of CFM-GP–predicted gene expression profiles for the same datasets, preserving the clustering patterns observed in the ground truth.
    (c) UMAP comparison of control vs. real perturbed cells (left) and control vs. predicted perturbed cells (right), showing that CFM-GP predictions closely match the ground-truth perturbation distribution.
}

    \label{fig:results_mmd_umap}
\end{figure}

Predicting gene perturbation responses across diverse cell types is challenging due to the highly heterogeneous and context-specific nature of transcriptional dynamics. Existing approaches often require separate models for each cell type or fail to explicitly incorporate cell identity, limiting their ability to generalize to unseen cellular contexts. CFM-GP addresses this gap by treating cell type as an explicit conditioning variable within a unified flow-matching framework, enabling a single model to learn shared perturbation principles while preserving cell-type–specific nuances.

In CFM-GP, the transformation from control to perturbed gene expression is formulated as a continuous trajectory in expression space, parameterized by a conditional neural vector field (Figure~\ref{fig:method_pipeline}). At each integration step, the vector field receives four inputs: the current interpolated expression state, the original control state, a learnable cell type embedding, and a continuous time embedding. This conditioning scheme allows the model to capture both the global, shared components of perturbation dynamics and the finer, cell-type–specific deviations that distinguish one population’s response from another. By integrating the learned vector field as an ordinary differential equation from \(t = 0\) (control) to \(t = 1\) (perturbed), CFM-GP produces predictions that remain coherent across diverse cell types while accurately reflecting the unique transcriptional shifts associated with each identity. This unified, cell type–aware formulation eliminates the need for training separate models per cell type, improves data efficiency by leveraging shared structure across populations, and facilitates generalization to contexts not seen during training.

The benefits of this design are evident in the UMAP visualizations (Figure~\ref{fig:results_mmd_umap}). Ground-truth profiles (Figure~\ref{fig:results_mmd_umap}a) display well-separated cell type clusters across diverse datasets, while CFM-GP predictions (Figure~\ref{fig:results_mmd_umap}b) faithfully preserve these boundaries, capturing both the intra-cell-type structure and inter-cell-type relationships. Furthermore, direct comparisons between control vs.\ real perturbed profiles and control vs.\ predicted perturbed profiles (Figure~\ref{fig:results_mmd_umap}c) reveal that the predicted perturbation distributions closely match the true ones, demonstrating that CFM-GP not only maintains overall population structure but also accurately reproduces biologically meaningful perturbation-induced shifts. These qualitative results highlight the model’s ability to unify learning across heterogeneous cell types without sacrificing the specificity of perturbation effects.

\subsection*{CFM-GP Demonstrates High Predictive Accuracy Across Five Datasets as Measured by R-Squared Value}

\begin{figure}[htbp]
    \centering
    \includegraphics[width=\textwidth]{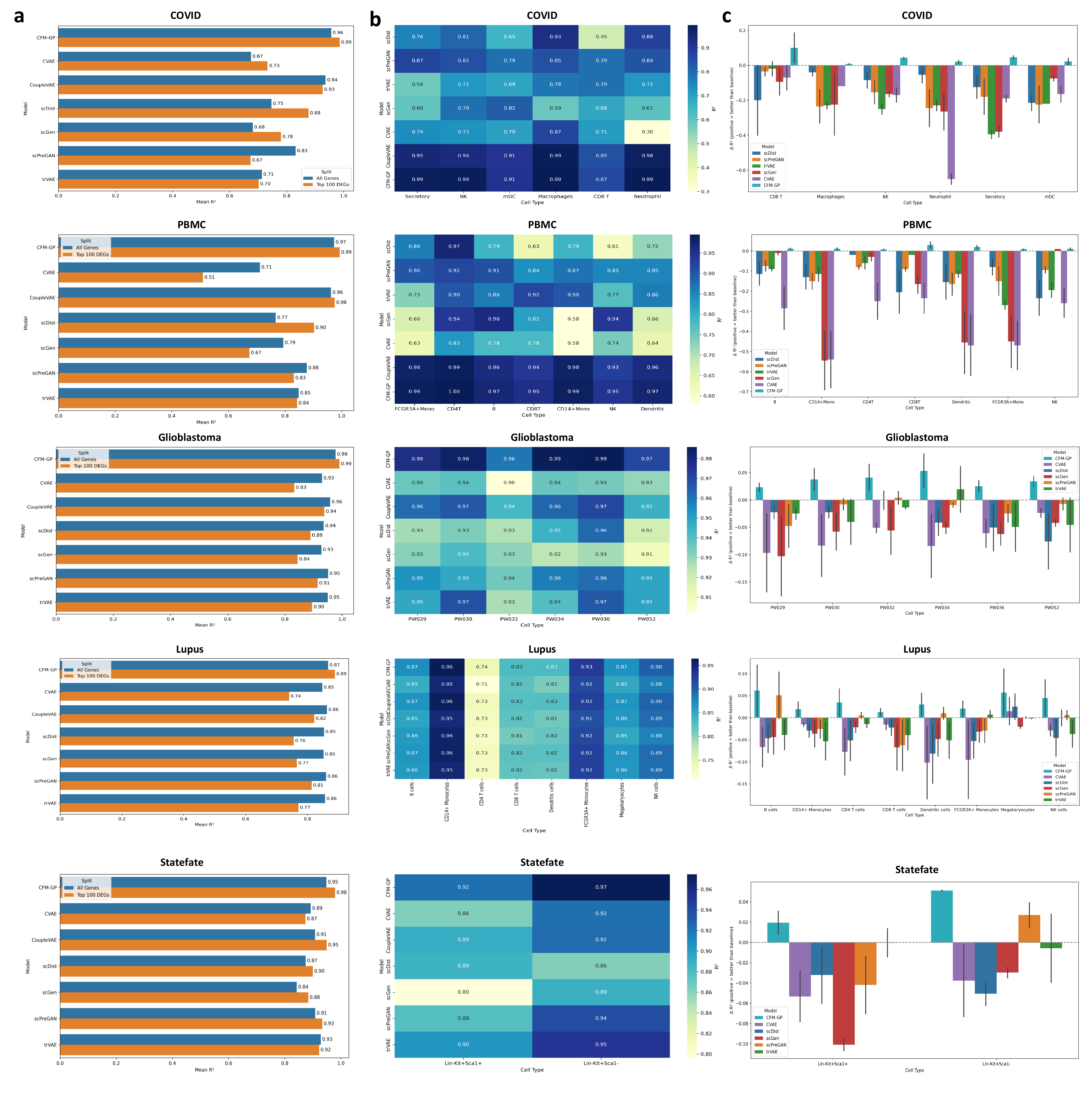}
    \caption{Comparison of predictive performance (R\textsuperscript{2}) across five datasets—COVID-19, PBMC, Glioblastoma, Lupus, and Statefate—between CFM-GP and existing baseline models. Panel (a) shows absolute R\textsuperscript{2} values per model and cell type. Panel (b) presents a heatmap of R\textsuperscript{2} values across all datasets and cell types.Panel (c) illustrates the performance improvement of CFM-GP relative to CoupleVAE (\(\Delta R^2\)). }

    \label{fig:results_r2}
\end{figure}

We evaluated the predictive accuracy of gene perturbation responses using R² values across multiple datasets, comparing our proposed method, CFM-GP, against the baseline method CoupleVAE, as well as benchmark models including scDist, scPreGAN, trVAE, scGen, and CVAE. R² scores, quantifying the proportion of variance in gene expression explained by the predictions, were calculated separately for the entire gene set ("All Genes") and the top 100 differentially expressed genes ("Top 100 DEGs").

The COVID dataset encompasses six distinct cell types: Secretory, NK, mDC, Macrophages, CD8 T, and Neutrophil. Across all these cell types, CFM-GP consistently demonstrated superior predictive accuracy compared to all benchmarking methods. For predictions involving all genes, CFM-GP achieved notably high R² scores, ranging from 0.865 in CD8 T cells to 0.995 in Macrophages shown in the bar plot (Figure \ref{fig:results_r2}a) and heatmap (Figure \ref{fig:results_r2}b). Relative to the baseline method CoupleVAE, CFM-GP showed consistent improvements across all cell types shown in the comparative bar plot (Figure \ref{fig:results_r2}c). The most substantial improvement was observed in NK cells, where CFM-GP achieved an R² of 0.987 compared to CoupleVAE's 0.94, reflecting a 5\% increase in predictive accuracy. The smallest observed improvement occurred in mDC cells, with a modest increase from 0.91 to 0.915 (+0.5\%). When analyzing only the top 100 DEGs, the performance gap widened considerably in favor of CFM-GP. R² scores surpassed 0.994 across all cell types, reflecting particularly pronounced predictive enhancements compared to CoupleVAE. For example, CD8 T cells exhibited the most significant improvement, where CFM-GP achieved an R² of 0.955, marking a 24\% increase over CoupleVAE's 0.77. Similarly, substantial gains were observed in Secretory cells (from 0.94 to 0.995, +5.8\%) and NK cells (from 0.96 to 0.997, +3.8\%). The smallest improvement in this subset was observed in Macrophages, where CoupleVAE's already high baseline of 0.99 was marginally improved to 0.999 (+0.9\%). Comparing CFM-GP with other benchmark methods (scDist, scPreGAN, trVAE, scGen, CVAE) further underscores its robust performance. For example, in CD8 T cells, CFM-GP significantly outperformed CVAE and scDist by 18.5\%, improving from approximately 0.77 to 0.955. In Macrophages, where all methods generally performed well, CFM-GP still notably enhanced predictive accuracy, raising the benchmark from approximately 0.97 achieved by scDist to a near-perfect R² score of 0.999.

\begin{table}[htbp]
\centering
\caption{R\textsuperscript{2} predictive accuracy across COVID-19 cell types for All Genes and Top 100 DEGs. Rows represent models; columns indicate cell types. Percentage improvements for CFM-GP over CoupleVAE are shown in the final rows.}
\label{tab:covid_r2_combined}
\resizebox{\textwidth}{!}{%
\begin{tabular}{lcccccc}
\toprule
\textbf{Model (Split)} & \textbf{Secretory} & \textbf{NK} & \textbf{mDC} & \textbf{Macrophages} & \textbf{CD8 T} & \textbf{Neutrophil} \\
\midrule
scDist (All Genes) & 0.76 & 0.81 & 0.65 & 0.93 & 0.45 & 0.88 \\
scPreGAN (All Genes) & 0.87 & 0.85 & 0.79 & 0.85 & 0.79 & 0.84 \\
trVAE (All Genes) & 0.58 & 0.72 & 0.69 & 0.78 & 0.79 & 0.72 \\
scGen (All Genes) & 0.60 & 0.79 & 0.82 & 0.59 & 0.68 & 0.61 \\
CVAE (All Genes) & 0.74 & 0.73 & 0.70 & 0.87 & 0.71 & 0.30 \\
CoupleVAE (All Genes) & 0.95 & 0.94 & 0.91 & 0.99 & 0.85 & 0.98 \\
\textbf{CFM-GP (All Genes)} & \textbf{0.9861} & \textbf{0.9875} & \textbf{0.9147} & \textbf{0.9946} & \textbf{0.8651} & \textbf{0.9928} \\
\midrule
scDist (Top 100 DEGs) & 0.88 & 0.92 & 0.76 & 0.97 & 0.77 & 0.96 \\
scPreGAN (Top 100 DEGs) & 0.66 & 0.74 & 0.60 & 0.66 & 0.76 & 0.62 \\
trVAE (Top 100 DEGs) & 0.52 & 0.68 & 0.71 & 0.74 & 0.79 & 0.77 \\
scGen (Top 100 DEGs) & 0.53 & 0.78 & 0.87 & 0.94 & 0.75 & 0.81 \\
CVAE (Top 100 DEGs) & 0.77 & 0.83 & 0.81 & 0.87 & 0.77 & 0.35 \\
CoupleVAE (Top 100 DEGs) & 0.94 & 0.96 & 0.93 & 0.99 & 0.77 & 0.97 \\
\textbf{CFM-GP (Top 100 DEGs)} & \textbf{0.9949} & \textbf{0.9967} & \textbf{0.9697} & \textbf{0.9991} & \textbf{0.9552} & \textbf{0.9992} \\
\midrule
\textit{\% Improvement (All Genes)} & +3.81\% & +5.05\% & +0.81\% & +0.46\% & +1.77\% & +1.30\% \\
\textit{\% Improvement (Top 100 DEGs)} & +5.88\% & +3.65\% & +4.27\% & +0.91\% & +24.06\% & +2.90\% \\
\bottomrule
\end{tabular}%
}
\end{table}

\begin{table}[htbp]
\centering
\caption{R\textsuperscript{2} predictive accuracy across PBMC cell types for All Genes and Top 100 DEGs. Rows represent models; columns indicate cell types. Percentage improvements for CFM-GP over CoupleVAE are shown in the final rows.}
\label{tab:pbmc_r2_combined}
\resizebox{\textwidth}{!}{%
\begin{tabular}{lccccccc}
\toprule
\textbf{Model (Split)} & \textbf{FCGR3A+Mono} & \textbf{CD4T} & \textbf{B} & \textbf{CD8T} & \textbf{CD14+Mono} & \textbf{NK} & \textbf{Dendritic} \\
\midrule
scDist (All Genes) & 0.86 & 0.97 & 0.79 & 0.63 & 0.79 & 0.61 & 0.72 \\
scPreGAN (All Genes) & 0.90 & 0.92 & 0.91 & 0.84 & 0.87 & 0.85 & 0.85 \\
trVAE (All Genes) & 0.73 & 0.90 & 0.86 & 0.92 & 0.90 & 0.77 & 0.86 \\
scGen (All Genes) & 0.66 & 0.94 & 0.96 & 0.82 & 0.58 & 0.94 & 0.66 \\
CVAE (All Genes) & 0.63 & 0.83 & 0.78 & 0.78 & 0.58 & 0.74 & 0.64 \\
CoupleVAE (All Genes) & 0.98 & 0.99 & 0.96 & 0.94 & 0.98 & 0.93 & 0.96 \\
\textbf{CFM-GP (All Genes)} & \textbf{0.9903} & \textbf{0.9955} & \textbf{0.9696} & \textbf{0.9537} & \textbf{0.9870} & \textbf{0.9457} & \textbf{0.9739} \\
\midrule
scDist (Top 100 DEGs) & 0.95 & 0.97 & 0.92 & 0.85 & 0.91 & 0.82 & 0.90 \\
scPreGAN (Top 100 DEGs) & 0.77 & 0.90 & 0.88 & 0.87 & 0.79 & 0.86 & 0.75 \\
trVAE (Top 100 DEGs) & 0.70 & 0.96 & 0.90 & 0.93 & 0.83 & 0.74 & 0.84 \\
scGen (Top 100 DEGs) & 0.41 & 0.98 & 0.96 & 0.74 & 0.29 & 0.98 & 0.36 \\
CVAE (Top 100 DEGs) & 0.40 & 0.65 & 0.59 & 0.64 & 0.30 & 0.64 & 0.35 \\
CoupleVAE (Top 100 DEGs) & 0.99 & 0.99 & 0.98 & 0.95 & 0.98 & 0.97 & 0.97 \\
\textbf{CFM-GP (Top 100 DEGs)} & \textbf{0.9982} & \textbf{0.9998} & \textbf{0.9941} & \textbf{0.9946} & \textbf{0.9954} & \textbf{0.9758} & \textbf{0.9966} \\
\midrule
\textit{\% Improvement (All Genes)} & +1.05\% & +0.55\% & +1.00\% & +1.46\% & +0.71\% & +1.69\% & +1.45\% \\
\textit{\% Improvement (Top 100 DEGs)} & +0.83\% & +0.99\% & +1.43\% & +4.71\% & +1.57\% & +0.60\% & +2.74\% \\
\bottomrule
\end{tabular}%
}
\end{table}

The PBMC dataset includes seven immune cell types: FCGR3A+ Monocytes, CD4 T cells, B cells, CD8 T cells, CD14+ Monocytes, NK cells, and Dendritic cells. CFM-GP again outperformed all benchmarking methods across every cell type. When using all genes, CFM-GP achieved R² scores ranging from 0.946 (NK) to 0.995 (CD4 T), surpassing CoupleVAE in all cases. The relative improvements in R² over CoupleVAE were modest but consistent, with the largest improvement observed in NK cells (+1.69\%) and the smallest in CD4 T cells (+0.55\%). The predictive gains became more substantial when focusing on the top 100 DEGs. CFM-GP reached near-perfect R² values across all cell types, including 0.999 for CD4 T and 0.998 for FCGR3A+ Monocytes. CD8 T cells saw the most pronounced improvement, increasing from 0.95 to 0.995 (+4.71\%). B cells and Dendritic cells also showed marked increases (+1.43\% and +2.74\%, respectively). Even in cell types where CoupleVAE performed well—such as FCGR3A+ Mono and CD14+ Mono—CFM-GP provided additional improvements. In contrast, models like CVAE and scGen frequently underperformed, particularly in monocyte populations, where R² values for top DEGs were as low as 0.30–0.40. 

In the Glioblastoma dataset, CFM-GP demonstrated improvements over CoupleVAE in every sample, with gains ranging from +1.59\% (PW036) to +2.66\% (PW052). The performance gap widened when evaluating the top 100 DEGs, where CFM-GP outperformed CoupleVAE by as much as +9.36\% in PW034. Notably, CFM-GP maintained an R² above 0.985 in all samples for the All Genes setting, while other methods—including trVAE and CVAE—showed variable performance, particularly in more heterogeneous samples such as PW032. 

In the lupus dataset, which comprises immune cell profiles from lupus patients under IFN-$\beta$ stimulation, CFM-GP consistently outperformed all baseline methods across eight evaluated cell types. In the all-genes setting, the improvements over CoupleVAE were modest but uniform, ranging from +0.40\% in NK cells to +0.74\% in CD8 T cells, indicating consistent generalization across inflammatory immune states. More pronounced gains were observed in the top 100 DEGs condition, where CFM-GP achieved the highest R² scores in all cell types. Notably, it delivered a +22.30\% improvement in Megakaryocytes—a population where most competing models, including CoupleVAE and CVAE, struggled to exceed R² scores of 0.55. CFM-GP also achieved near-perfect predictive accuracy in CD14+ Monocytes (R² = 0.994) and FCGR3A+ Monocytes (R² = 0.984), outperforming both deep generative and deterministic baselines.

Despite involving only two progenitor subpopulations, the Statefate data revealed nuanced differences in how models captured cytokine-induced transcriptional shifts. CFM-GP stood out for its ability to generalize across both early stem-like (\textit{Lin-Kit+Sca1+}) and more committed (\textit{Lin-Kit+Sca1-}) cells. In the all-genes setting, it delivered improvements of +3.5\% and +5.5\% R\textsuperscript{2} over CoupleVAE, with further refinement observed in the top 100 DEGs condition. Notably, the Sca1- population---often more transcriptionally variable---saw the greatest benefit, with CFM-GP achieving near-perfect R\textsuperscript{2} (0.996), outperforming all baselines. 


\begin{table}[htbp]
\centering
\caption{R\textsuperscript{2} predictive accuracy across glioblastoma samples for All Genes and Top 100 DEGs. Rows represent models; columns indicate sample IDs. Percentage improvements for CFM-GP over CoupleVAE are shown in the final rows.}
\label{tab:glioblastoma_r2_combined}
\resizebox{0.8\textwidth}{!}{
\begin{tabular}{lcccccc}
\toprule
\textbf{Model (Split)} & \textbf{PW029} & \textbf{PW030} & \textbf{PW032} & \textbf{PW034} & \textbf{PW036} & \textbf{PW052} \\
\midrule
CoupleVAE (All Genes) & 0.962 & 0.966 & 0.943 & 0.963 & 0.970 & 0.945 \\
scPreGAN (All Genes) & 0.953 & 0.948 & 0.935 & 0.958 & 0.962 & 0.949 \\
trVAE (All Genes) & 0.948 & 0.967 & 0.927 & 0.941 & 0.966 & 0.949 \\
scDist (All Genes) & 0.928 & 0.933 & 0.926 & 0.945 & 0.963 & 0.919 \\
scGen (All Genes) & 0.932 & 0.942 & 0.931 & 0.924 & 0.928 & 0.909 \\
CVAE (All Genes) & 0.936 & 0.939 & 0.901 & 0.937 & 0.935 & 0.930 \\
\textbf{CFM-GP (All Genes)} & \textbf{0.978} & \textbf{0.983} & \textbf{0.960} & \textbf{0.985} & \textbf{0.986} & \textbf{0.970} \\
\midrule
CoupleVAE (Top 100 DEGs) & 0.957 & 0.939 & 0.926 & 0.904 & 0.954 & 0.948 \\
scPreGAN (Top 100 DEGs) & 0.870 & 0.941 & 0.942 & 0.891 & 0.914 & 0.930 \\
trVAE (Top 100 DEGs) & 0.921 & 0.857 & 0.915 & 0.966 & 0.860 & 0.853 \\
scDist (Top 100 DEGs) & 0.946 & 0.927 & 0.942 & 0.840 & 0.861 & 0.822 \\
scGen (Top 100 DEGs) & 0.781 & 0.847 & 0.827 & 0.843 & 0.871 & 0.900 \\
CVAE (Top 100 DEGs) & 0.788 & 0.798 & 0.866 & 0.762 & 0.867 & 0.917 \\
\textbf{CFM-GP (Top 100 DEGs)} & \textbf{0.988} & \textbf{0.996} & \textbf{0.992} & \textbf{0.989} & \textbf{0.989} & \textbf{0.991} \\
\midrule
\textit{\% Improvement (All Genes)} & +1.66 & +1.79 & +1.76 & +2.31 & +1.59 & +2.66 \\
\textit{\% Improvement (Top 100 DEGs)} & +3.23 & +6.04 & +7.11 & +9.36 & +3.68 & +4.58 \\
\bottomrule
\end{tabular}}
\end{table}

\begin{table}[htbp]
\centering
\caption{R\textsuperscript{2} predictive accuracy across lupus PBMC cell types for All Genes and Top 100 DEGs. Rows represent models; columns indicate cell types. Percentage improvements for CFM-GP over CoupleVAE are shown in the final rows.}
\label{tab:lupus_r2_combined}
\resizebox{\textwidth}{!}{%
\begin{tabular}{lcccccccc}
\toprule
\textbf{Model (Split)} & \textbf{B cells} & \textbf{CD14+ Mono} & \textbf{CD4 T} & \textbf{CD8 T} & \textbf{Dendritic} & \textbf{FCGR3A+ Mono} & \textbf{Megakaryocytes} & \textbf{NK cells} \\
\midrule
CoupleVAE (All Genes) & 0.867 & 0.961 & 0.735 & 0.826 & 0.828 & 0.925 & 0.866 & 0.898 \\
scPreGAN (All Genes) & 0.865 & 0.957 & 0.733 & 0.823 & 0.825 & 0.923 & 0.864 & 0.893 \\
trVAE (All Genes) & 0.862 & 0.950 & 0.729 & 0.821 & 0.818 & 0.923 & 0.864 & 0.890 \\
scDist (All Genes) & 0.852 & 0.947 & 0.729 & 0.822 & 0.813 & 0.911 & 0.862 & 0.893 \\
scGen (All Genes) & 0.862 & 0.955 & 0.726 & 0.811 & 0.818 & 0.918 & 0.849 & 0.879 \\
CVAE (All Genes) & 0.846 & 0.951 & 0.710 & 0.823 & 0.808 & 0.917 & 0.849 & 0.881 \\
\textbf{CFM-GP (All Genes)} & \textbf{0.871} & \textbf{0.965} & \textbf{0.738} & \textbf{0.831} & \textbf{0.832} & \textbf{0.929} & \textbf{0.869} & \textbf{0.901} \\
\midrule
CoupleVAE (Top 100 DEGs) & 0.812 & 0.958 & 0.783 & 0.807 & 0.887 & 0.946 & 0.497 & 0.887 \\
scPreGAN (Top 100 DEGs) & 0.915 & 0.911 & 0.795 & 0.686 & 0.911 & 0.889 & 0.499 & 0.903 \\
trVAE (Top 100 DEGs) & 0.739 & 0.862 & 0.761 & 0.734 & 0.795 & 0.962 & 0.496 & 0.820 \\
scDist (Top 100 DEGs) & 0.733 & 0.914 & 0.686 & 0.766 & 0.738 & 0.854 & 0.551 & 0.800 \\
scGen (Top 100 DEGs) & 0.729 & 0.891 & 0.749 & 0.688 & 0.800 & 0.891 & 0.473 & 0.905 \\
CVAE (Top 100 DEGs) & 0.700 & 0.938 & 0.653 & 0.780 & 0.703 & 0.763 & 0.543 & 0.846 \\
\textbf{CFM-GP (Top 100 DEGs)} & \textbf{0.932} & \textbf{0.994} & \textbf{0.849} & \textbf{0.828} & \textbf{0.942} & \textbf{0.984} & \textbf{0.608} & \textbf{0.973} \\
\midrule
\textit{\% Improvement (All Genes)} & +0.45\% & +0.45\% & +0.41\% & +0.74\% & +0.54\% & +0.43\% & +0.41\% & +0.40\% \\
\textit{\% Improvement (Top 100 DEGs)} & +14.79\% & +3.69\% & +8.42\% & +2.64\% & +6.13\% & +4.06\% & +22.30\% & +9.71\% \\
\bottomrule
\end{tabular}%
}
\end{table}

\begin{table}[htbp]
\centering
\caption{R\textsuperscript{2} predictive accuracy across Statefate progenitor populations for All Genes and Top 100 DEGs. Rows represent models; columns indicate sorted cell populations. Percentage improvements for CFM-GP over CoupleVAE are shown in the final rows.}
\label{tab:statefate_r2_combined}
\resizebox{0.5\textwidth}{!}{%
\begin{tabular}{lcc}
\toprule
\textbf{Model (Split)} & \textbf{Lin-Kit+Sca1+} & \textbf{Lin-Kit+Sca1-} \\
\midrule
CoupleVAE (All Genes) & 0.891 & 0.924 \\
scPreGAN (All Genes) & 0.878 & 0.939 \\
trVAE (All Genes) & 0.905 & 0.952 \\
scDist (All Genes) & 0.887 & 0.862 \\
scGen (All Genes) & 0.796 & 0.889 \\
CVAE (All Genes) & 0.863 & 0.922 \\
\textbf{CFM-GP (All Genes)} & \textbf{0.922} & \textbf{0.975} \\
\midrule
CoupleVAE (Top 100 DEGs) & 0.954 & 0.945 \\
scPreGAN (Top 100 DEGs) & 0.884 & 0.984 \\
trVAE (Top 100 DEGs) & 0.940 & 0.905 \\
scDist (Top 100 DEGs) & 0.894 & 0.906 \\
scGen (Top 100 DEGs) & 0.847 & 0.920 \\
CVAE (Top 100 DEGs) & 0.876 & 0.871 \\
\textbf{CFM-GP (Top 100 DEGs)} & \textbf{0.962} & \textbf{0.996} \\
\midrule
\textit{\% Improvement (All Genes)} & +3.49\% & +5.50\% \\
\textit{\% Improvement (Top 100 DEGs)} & +0.84\% & +5.46\% \\
\bottomrule
\end{tabular}%
}
\end{table}

\begin{figure}[htbp]
    \centering
    \includegraphics[width=0.9\textwidth]{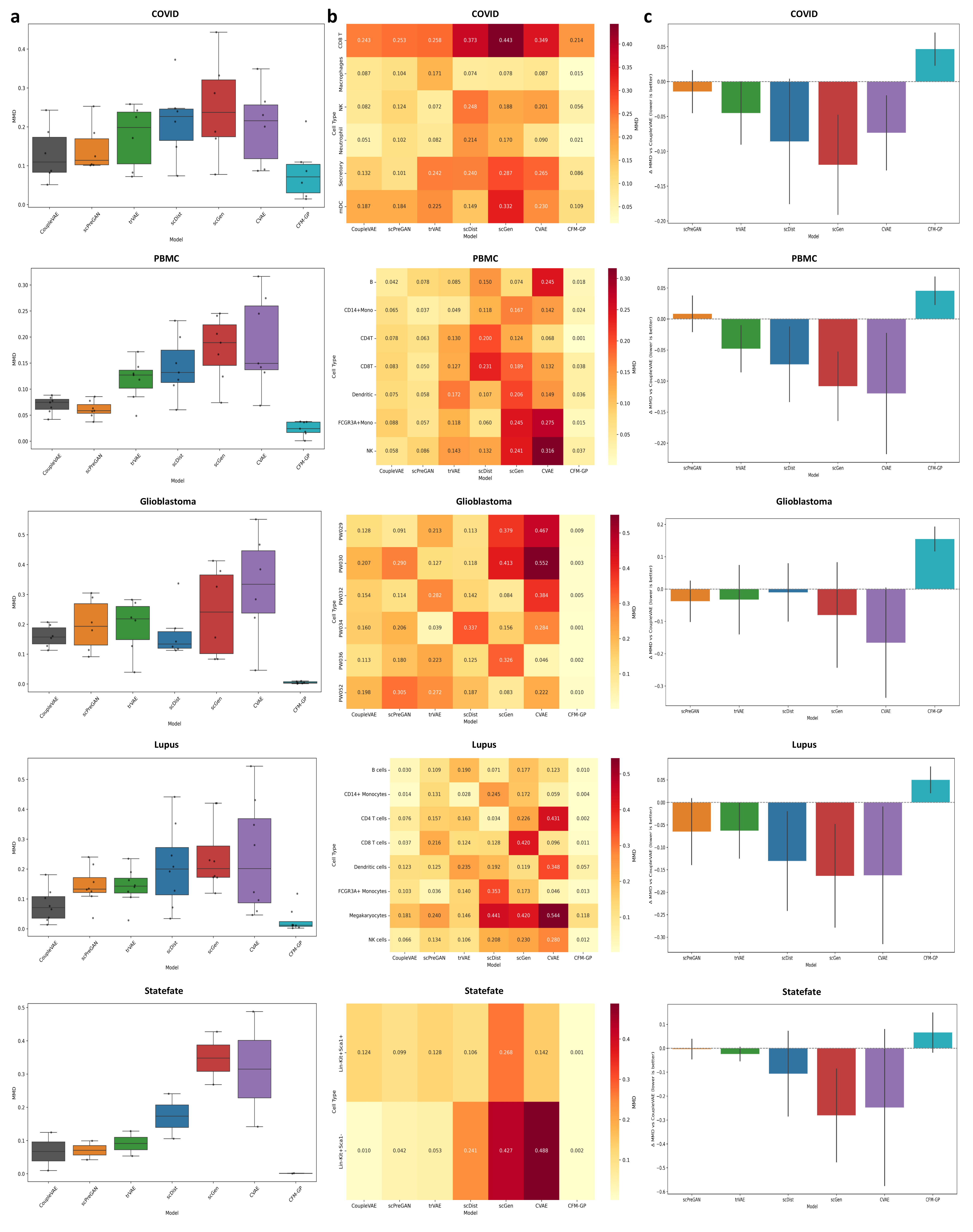}
    \caption{Comparison of distributional similarity between predicted and real gene expression across models using Maximum Mean Discrepancy (MMD). 
    \textbf{(a)} Boxplots showing MMD values for each model across five datasets (COVID, PBMC, Glioblastoma, Lupus, Statefate); lower values indicate better alignment. 
    \textbf{(b)} Heatmaps of MMD scores per model and cell type, highlighting CFM-GP’s consistent performance across diverse biological settings. 
    \textbf{(c)} Bar plots showing average MMD improvement (\(\Delta\)MMD) of each model relative to CoupleVAE, with CFM-GP consistently achieving the largest reductions in distributional divergence.}

    \label{fig:results_mmd}
\end{figure}

\subsection*{CFM-GP Accurately Captures Distributional Similarity with Low Maximum Mean Discrepancy (MMD) Across Diverse Datasets}

In addition to gene-level accuracy metrics, we evaluated how closely the predicted gene expression distributions matched the real perturbed distributions using Maximum Mean Discrepancy (MMD). As a kernel-based statistical measure of distributional distance, MMD provides a complementary assessment to pointwise metrics such as R\textsuperscript{2}. Lower MMD values indicate better alignment between the predicted and observed perturbation distributions, reflecting the biological plausibility of a model's generative capacity.

Across all six COVID-19 cell types, CFM-GP achieved the lowest MMD scores compared to all baseline methods (Figure~\ref{fig:results_mmd}a, Table~\ref{tab:covid_mmd_results}). For example, CFM-GP reduced MMD by over 83\% relative to CoupleVAE in macrophages and by 58\% in neutrophils. Improvements were also notable in NK cells (31\%) and secretory cells (35\%), where alternative models such as CVAE and scGen showed substantially higher divergence. Although trVAE slightly outperformed CoupleVAE in NK cells, CFM-GP remained the most consistent performer across all subsets. 

The PBMC dataset further demonstrates CFM-GP’s strength in capturing accurate perturbation distributions across immune subtypes. As shown in Table~\ref{tab:pbmc_mmd_results} and Figure~\ref{fig:results_mmd}a, CFM-GP achieved the lowest MMD values in all seven evaluated cell types. In B cells and CD8 T cells, MMD was reduced by over 50\% relative to CoupleVAE. The largest reduction was observed in CD4 T cells, where CFM-GP nearly eliminated distributional divergence (MMD = 0.0008, +98.9\% improvement). Heatmaps (Figure~\ref{fig:results_mmd}b) highlight broad divergence in baseline methods, particularly CVAE and scGen, whereas CFM-GP maintained a narrow and consistent distribution across all subsets. Additionally, the delta-MMD plot (Figure~\ref{fig:results_mmd}c) shows that CFM-GP consistently yields lower MMD than CoupleVAE, unlike several other models that show cell-type–dependent variation or even worse performance. 

Distributional evaluation on glioblastoma patient samples confirmed the robustness of CFM-GP across individual-level heterogeneity. As detailed in Table~\ref{tab:glioblastoma_mmd_results}, CFM-GP achieved the lowest MMD scores across all six samples, with reductions ranging from +93\% to +99\% over CoupleVAE. In PW034, the model reduced the MMD from 0.160 to 0.0009, a 99.5\% improvement, while even more modest samples like PW029 showed over 93\% reduction. Compared to competing methods, including trVAE, CVAE, and scGen, which exhibited substantial variance and outlier behavior (Figure~\ref{fig:results_mmd}a), CFM-GP maintained a tightly clustered low-discrepancy range. Heatmaps (Figure~\ref{fig:results_mmd}b) further illustrated how CFM-GP uniformly minimized distributional mismatch across all patients, and delta-MMD plots (Figure~\ref{fig:results_mmd}c) reinforced that these gains were not driven by a single sample but generalized across the cohort.

In the lupus dataset, which captures IFN-$\beta$--stimulated responses in autoimmune patient-derived immune cells, CFM-GP again demonstrated the strongest distributional alignment. According to Table~\ref{tab:lupus_mmd_results} and Figure~\ref{fig:results_mmd}a, CFM-GP consistently achieved the lowest MMD values across all eight cell types. Reductions relative to CoupleVAE were substantial: +97\% for CD4 T cells, +87\% for FCGR3A+ monocytes, and over +81\% for NK cells. Even in more difficult populations such as megakaryocytes, where all models struggled, CFM-GP offered the lowest discrepancy (MMD = 0.1176), outperforming others by a wide margin. The heatmap in Figure~\ref{fig:results_mmd}b clearly visualizes this trend, with CFM-GP's column standing out as uniformly light across rows, in contrast to the darker, more variable entries of baseline methods. The delta-MMD summary (Figure~\ref{fig:results_mmd}c) confirms CFM-GP’s broad and consistent advantage, underscoring its capacity to learn perturbation distributions faithfully even in immunologically dysregulated conditions.

Though composed of only two progenitor compartments, the Statefate dataset revealed sharp contrasts in model fidelity. CFM-GP achieved the lowest MMD in both Lin-Kit+Sca1+ and Lin-Kit+Sca1- populations, with near-complete alignment in the former (MMD = 0.0007, +99.5\% over CoupleVAE). Figure~\ref{fig:results_mmd}a highlights this striking contrast: while most models---including trVAE and scDist---struggled with distributional consistency, CFM-GP’s predictions were tightly concentrated and well-aligned. The corresponding delta-MMD plot (Figure~\ref{fig:results_mmd}b) shows CFM-GP as the only model yielding consistent positive gains in both subsets. These findings reinforce CFM-GP’s robustness not only in immune perturbation settings but also in developmental transitions involving subtle transcriptional shifts.

\begin{table}[htbp]
\centering
\caption{Maximum Mean Discrepancy (MMD) between predicted and real gene expression distributions for COVID-19 cell types across models. Lower values indicate better alignment. The rightmost column shows the improvement of CFM-GP over the baseline CoupleVAE.}
\label{tab:covid_mmd_results}
\begin{tabular}{lccccccc|c}
\toprule
\textbf{Cell Type} & \textbf{CoupleVAE} & \textbf{scPreGAN} & \textbf{trVAE} & \textbf{scDist} & \textbf{scGen} & \textbf{CVAE} & \textbf{CFM-GP} & \textbf{Improvement} \\
\midrule
CD8 T & 0.243 & 0.253 & 0.258 & 0.373 & 0.443 & 0.349 & \textbf{0.214} & +11.82\% \\
Macrophages & 0.087 & 0.104 & 0.171 & 0.074 & 0.078 & 0.087 & \textbf{0.015} & +83.29\% \\
NK & 0.082 & 0.124 & 0.072 & 0.248 & 0.188 & 0.201 & \textbf{0.056} & +31.26\% \\
Neutrophil & 0.051 & 0.102 & 0.082 & 0.214 & 0.170 & 0.090 & \textbf{0.021} & +58.27\% \\
Secretory & 0.132 & 0.101 & 0.242 & 0.240 & 0.287 & 0.265 & \textbf{0.086} & +34.91\% \\
mDC & 0.187 & 0.184 & 0.225 & 0.149 & 0.332 & 0.230 & \textbf{0.109} & +41.46\% \\
\bottomrule
\end{tabular}
\end{table}

\begin{table}[htbp]
\centering
\caption{Maximum Mean Discrepancy (MMD) between predicted and real gene expression distributions for PBMC cell types across models. Lower values indicate better alignment. The rightmost column shows the percentage improvement of CFM-GP over CoupleVAE.}
\label{tab:pbmc_mmd_results}
\begin{tabular}{lccccccc|c}
\toprule
\textbf{Cell Type} & \textbf{CoupleVAE} & \textbf{scPreGAN} & \textbf{trVAE} & \textbf{scDist} & \textbf{scGen} & \textbf{CVAE} & \textbf{CFM-GP} & \textbf{Improvement} \\
\midrule
B & 0.0419 & 0.0776 & 0.0852 & 0.1498 & 0.0740 & 0.2450 & \textbf{0.0182} & +56.61\% \\
CD14+Mono & 0.0647 & 0.0370 & 0.0486 & 0.1181 & 0.1668 & 0.1421 & \textbf{0.0241} & +62.79\% \\
CD4T & 0.0775 & 0.0631 & 0.1299 & 0.1999 & 0.1242 & 0.0683 & \textbf{0.0008} & +98.94\% \\
CD8T & 0.0834 & 0.0496 & 0.1269 & 0.2314 & 0.1892 & 0.1321 & \textbf{0.0377} & +54.74\% \\
Dendritic & 0.0746 & 0.0585 & 0.1717 & 0.1073 & 0.2062 & 0.1494 & \textbf{0.0364} & +51.24\% \\
FCGR3A+Mono & 0.0883 & 0.0575 & 0.1181 & 0.0603 & 0.2453 & 0.2746 & \textbf{0.0152} & +82.84\% \\
NK & 0.0575 & 0.0856 & 0.1428 & 0.1320 & 0.2408 & 0.3164 & \textbf{0.0370} & +35.68\% \\
\bottomrule
\end{tabular}
\end{table}

\begin{table}[htbp]
\centering
\caption{Maximum Mean Discrepancy (MMD) between predicted and real gene expression distributions for glioblastoma samples across models. Lower values indicate better alignment. The rightmost column shows the percentage improvement of CFM-GP over CoupleVAE.}
\label{tab:glioblastoma_mmd_results}
\begin{tabular}{lccccccc|c}
\toprule
\textbf{Sample ID} & \textbf{CoupleVAE} & \textbf{scPreGAN} & \textbf{trVAE} & \textbf{scDist} & \textbf{scGen} & \textbf{CVAE} & \textbf{CFM-GP} & \textbf{Improvement} \\
\midrule
PW029 & 0.128 & 0.091 & 0.213 & 0.113 & 0.379 & 0.467 & \textbf{0.0085} & +93.39\% \\
PW030 & 0.207 & 0.290 & 0.127 & 0.118 & 0.413 & 0.552 & \textbf{0.0035} & +98.33\% \\
PW032 & 0.154 & 0.114 & 0.282 & 0.142 & 0.084 & 0.384 & \textbf{0.0051} & +96.72\% \\
PW034 & 0.160 & 0.206 & 0.039 & 0.337 & 0.156 & 0.284 & \textbf{0.0009} & +99.45\% \\
PW036 & 0.113 & 0.180 & 0.223 & 0.125 & 0.326 & 0.0459 & \textbf{0.0022} & +98.09\% \\
PW052 & 0.198 & 0.305 & 0.272 & 0.187 & 0.0832 & 0.222 & \textbf{0.0100} & +94.97\% \\
\bottomrule
\end{tabular}
\end{table}

\begin{table}[htbp]
\centering
\caption{Maximum Mean Discrepancy (MMD) between predicted and real gene expression distributions for lupus cell types across models. Lower values indicate better alignment. The rightmost column shows the percentage improvement of CFM-GP over CoupleVAE.}
\label{tab:lupus_mmd_results}
\begin{tabular}{lccccccc|c}
\toprule
\textbf{Cell Type} & \textbf{CoupleVAE} & \textbf{scPreGAN} & \textbf{trVAE} & \textbf{scDist} & \textbf{scGen} & \textbf{CVAE} & \textbf{CFM-GP} & \textbf{Improvement} \\
\midrule
B cells & 0.0299 & 0.1092 & 0.1898 & 0.0715 & 0.1768 & 0.1227 & \textbf{0.0095} & +68.11\% \\
CD14+ Mono & 0.0138 & 0.1314 & 0.0284 & 0.2454 & 0.1718 & 0.0594 & \textbf{0.0043} & +68.82\% \\
CD4 T cells & 0.0764 & 0.1566 & 0.1627 & 0.0344 & 0.2261 & 0.4308 & \textbf{0.0023} & +96.95\% \\
CD8 T cells & 0.0374 & 0.2158 & 0.1245 & 0.1278 & 0.4205 & 0.0961 & \textbf{0.0114} & +69.61\% \\
Dendritic cells & 0.1226 & 0.1255 & 0.2347 & 0.1921 & 0.1191 & 0.3478 & \textbf{0.0571} & +53.39\% \\
FCGR3A+ Mono & 0.1032 & 0.0363 & 0.1400 & 0.3527 & 0.1726 & 0.0463 & \textbf{0.0134} & +87.00\% \\
Megakaryocytes & 0.1811 & 0.2401 & 0.1456 & 0.4415 & 0.4202 & 0.5440 & \textbf{0.1176} & +35.07\% \\
NK cells & 0.0656 & 0.1345 & 0.1059 & 0.2079 & 0.2295 & 0.2802 & \textbf{0.0124} & +81.07\% \\
\bottomrule
\end{tabular}
\end{table}

\begin{table}[htbp]
\centering
\caption{Maximum Mean Discrepancy (MMD) between predicted and real gene expression distributions for Statefate progenitor populations across models. Lower values indicate better alignment. The rightmost column shows the percentage improvement of CFM-GP over CoupleVAE.}
\label{tab:statefate_mmd_results}
\begin{tabular}{lccccccc|c}
\toprule
\textbf{Cell Type} & \textbf{CoupleVAE} & \textbf{scPreGAN} & \textbf{trVAE} & \textbf{scDist} & \textbf{scGen} & \textbf{CVAE} & \textbf{CFM-GP} & \textbf{Improvement} \\
\midrule
Lin-Kit+Sca1+ & 0.1245 & 0.0989 & 0.1284 & 0.1056 & 0.2682 & 0.1417 & \textbf{0.0007} & +99.48\% \\
Lin-Kit+Sca1- & 0.0096 & 0.0421 & 0.0534 & 0.2408 & 0.4274 & 0.4879 & \textbf{0.0017} & +82.37\% \\
\bottomrule
\end{tabular}
\end{table}

\subsection*{CFM-GP Preserves Gene Expression Ordering Across Multiple Datasets}
\begin{figure}[htbp]
    \centering
    \includegraphics[width=0.92\textwidth]{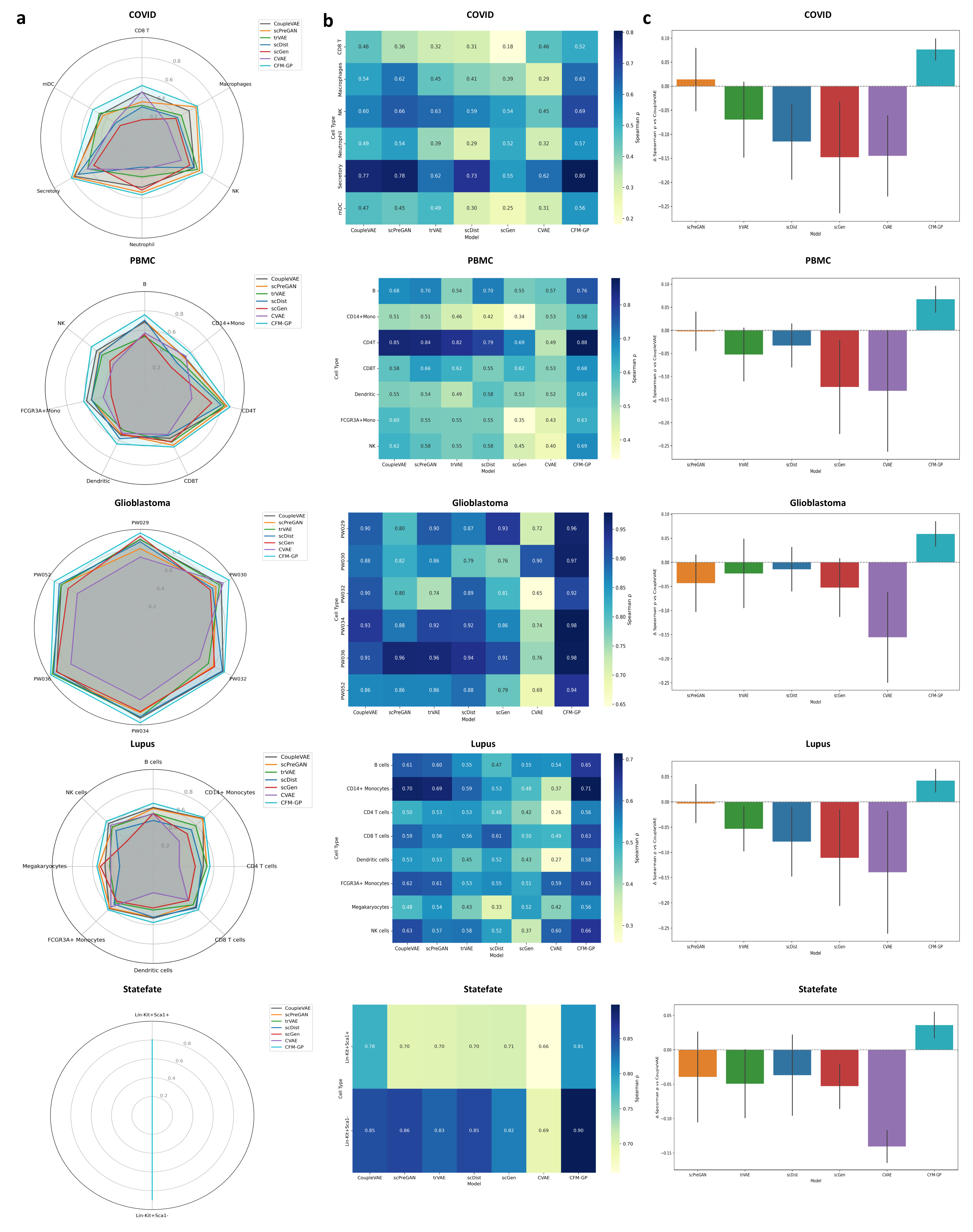}
    \caption{\textbf{Evaluation of gene ranking preservation using Spearman correlation.}
    \textbf{(a)} Radar plots showing average Spearman correlation (\(\rho\)) across cell types for each model. 
    \textbf{(b)} Heatmap of Spearman \(\rho\) values across models and cell types, where higher values indicate better consistency in predicted gene expression rankings.
    \textbf{(c)} Mean improvement in Spearman \(\rho\) of each model over the baseline (CoupleVAE), evaluated using paired t-tests across cell types. }

    \label{fig:results_spearman}
\end{figure}

To evaluate whether perturbation models retain the relative structure of gene expression changes, we computed the Spearman rank correlation (\(\rho\)) between predicted and real gene expression profiles. Unlike R\textsuperscript{2}, which assesses absolute correspondence, Spearman \(\rho\) captures the monotonic relationship between predicted and ground truth gene rankings—an important feature when assessing biological relevance in differential expression and pathway inference tasks.

We benchmarked CFM-GP against six competitive models across all five datasets using per-cell-type Spearman correlations. Figures~\ref{fig:results_spearman} visualize the performance: radar plots summarize per-model consistency across cell types; heatmaps reveal localized performance variation; and delta plots illustrate each model's gain over the baseline method (CoupleVAE). 

As shown in Table~\ref{tab:covid_spearman_results}, CFM-GP consistently achieved the highest Spearman \(\rho\) values across all six cell types, indicating that it most accurately preserved the rank ordering of gene expression following perturbation. Compared to the strongest baseline, CoupleVAE, CFM-GP improved Spearman correlation by +14.7\% in CD8 T cells, +17.4\% in macrophages, and nearly +20\% in mDCs (Table \ref{tab:covid_spearman_results}). Notably, even in cell types like NK and neutrophils where baselines such as scPreGAN and scGen performed competitively, CFM-GP still delivered the top ranking fidelity. These trends are evident in the radar view (Figure~\ref{fig:results_spearman}a) and heatmap (Figure~\ref{fig:results_spearman}b ), where CFM-GP shows uniform superiority. The bar chart of delta-Spearman (Figure~\ref{fig:results_spearman}c) further confirms that CFM-GP provides consistent rank-preserving benefits across cellular contexts.

In the PBMC dataset, CFM-GP outperforms all competing models across all seven cell types, achieving the highest Spearman correlation in each case shown in Table \ref{tab:pbmc_spearman_results}. While CoupleVAE generally ranked second—particularly in CD4T and FCGR3A+ monocytes—its performance lagged further behind in dendritic and CD8T cells, where models like scPreGAN and scDist came closer. The largest relative improvements from CFM-GP were observed in dendritic cells (+18.0\%) and CD8T cells (+17.1\%), suggesting enhanced sensitivity to subtle ordering shifts in more heterogeneous or transcriptionally variable populations. CD4T cells saw the smallest margin of improvement (+3.9\%), possibly due to already high baseline performance from CoupleVAE in this subset. Across models, trVAE and CVAE consistently ranked lowest, often underperforming even scDist, while scPreGAN performed competitively only in B cells.

As shown in Table~\ref{tab:glioblastoma_spearman_results}, CFM-GP achieved the highest Spearman rank correlation in all six glioblastoma patient samples. The strongest relative gains were observed in PW029 (+7.5\%), PW030 (+9.5\%), and PW052 (+8.7\%), where CFM-GP consistently outperformed both the baseline CoupleVAE and all other methods. Interestingly, while scGen exhibited competitive performance in PW029 (0.933) and trVAE matched or exceeded CoupleVAE in several instances (e.g., PW036), these methods exhibited more variability across patients. For instance, CVAE and scPreGAN had lower and less consistent performance, often trailing by a wide margin. In PW032 and PW034, the improvement of CFM-GP over CoupleVAE was more modest (+1.5\% and +5.0\%, respectively), likely due to already strong baseline performance.

The lupus dataset, derived from interferon-$\beta$–stimulated patient samples, exhibited broader variability in Spearman correlations across models and cell types. CFM-GP delivered top performance for all eight populations, with the greatest relative gains in megakaryocytes (+16.5\%) and CD4 T cells (+13.3\%) shown in Table \ref{tab:lupus_spearman_results}. These improvements suggest that CFM-GP is particularly robust in transcriptionally noisy or heterogeneous compartments. On the other hand, baseline models like CoupleVAE performed reasonably well in CD14+ monocytes and FCGR3A+ monocytes, narrowing CFM-GP’s margin to under 3\%. While trVAE and scPreGAN occasionally came close, such as in B and CD4 T cells, neither consistently ranked second. CVAE and scDist underperformed in nearly all populations. 

In the Statefate dataset, which focuses on two hematopoietic progenitor populations, CFM-GP achieved the highest Spearman rank correlation in both Lin-Kit+Sca1+ and Lin-Kit+Sca1- subsets, with modest but consistent gains over the best baselines. While CoupleVAE and scPreGAN delivered relatively strong performance—especially in Lin-Kit+Sca1- CFM-GP still surpassed them, improving correlation by +2.8\% and +5.8\% respectively (Table \ref{tab:statefate_spearman_results}). Interestingly, while differences across models were smaller than in other datasets, CFM-GP's edge was most apparent in the more transcriptionally complex Lin-Kit+Sca1- population, hinting at its sensitivity to nuanced gene ordering. CVAE trailed significantly in both cell types, and scDist underperformed in Lin-Kit+Sca1+, reaffirming CFM-GP’s ability to preserve rank structure in subtle developmental contexts.

\begin{table}[htbp]
\centering
\caption{Spearman rank correlation (\(\rho\)) between predicted and actual gene expression profiles for COVID-19 dataset. Higher values indicate better preservation of gene expression rankings. Rightmost column reports the improvement of CFM-GP over CoupleVAE.}
\label{tab:covid_spearman_results}
\begin{tabular}{lccccccc|c}
\toprule
\textbf{Cell Type} & \textbf{CoupleVAE} & \textbf{scPreGAN} & \textbf{trVAE} & \textbf{scDist} & \textbf{scGen} & \textbf{CVAE} & \textbf{CFM-GP} & \textbf{Improvement (\%)} \\
\midrule
CD8 T & 0.455 & 0.360 & 0.324 & 0.308 & 0.181 & 0.460 & \textbf{0.522} & +14.7\% \\
Macrophages & 0.538 & 0.617 & 0.449 & 0.405 & 0.387 & 0.288 & \textbf{0.632} & +17.4\% \\
NK & 0.596 & 0.656 & 0.633 & 0.594 & 0.545 & 0.449 & \textbf{0.689} & +15.6\% \\
Neutrophil & 0.494 & 0.544 & 0.389 & 0.293 & 0.521 & 0.316 & \textbf{0.570} & +15.4\% \\
Secretory & 0.767 & 0.782 & 0.620 & 0.729 & 0.550 & 0.623 & \textbf{0.805} & +5.0\% \\
mDC & 0.469 & 0.445 & 0.486 & 0.299 & 0.247 & 0.314 & \textbf{0.561} & +19.7\% \\
\bottomrule
\end{tabular}
\end{table}

\begin{table}[htbp]
\centering
\caption{Spearman rank correlation (\(\rho\)) between predicted and actual gene expression profiles for PBMC dataset. Higher values indicate better preservation of gene expression rankings. Rightmost column reports the improvement of CFM-GP over CoupleVAE.}
\label{tab:pbmc_spearman_results}
\begin{tabular}{lccccccc|c}
\toprule
\textbf{Cell Type} & \textbf{CoupleVAE} & \textbf{scPreGAN} & \textbf{trVAE} & \textbf{scDist} & \textbf{scGen} & \textbf{CVAE} & \textbf{CFM-GP} & \textbf{Improvement (\%)} \\
\midrule
B & 0.684 & 0.698 & 0.535 & 0.703 & 0.552 & 0.571 & \textbf{0.760} & +11.1\% \\
CD14+Mono & 0.515 & 0.507 & 0.458 & 0.416 & 0.343 & 0.530 & \textbf{0.584} & +13.5\% \\
CD4T & 0.846 & 0.842 & 0.820 & 0.787 & 0.691 & 0.486 & \textbf{0.879} & +3.9\% \\
CD8T & 0.579 & 0.658 & 0.622 & 0.547 & 0.618 & 0.534 & \textbf{0.678} & +17.1\% \\
Dendritic & 0.546 & 0.535 & 0.488 & 0.583 & 0.532 & 0.521 & \textbf{0.644} & +18.0\% \\
FCGR3A+Mono & 0.599 & 0.551 & 0.549 & 0.548 & 0.346 & 0.428 & \textbf{0.628} & +4.8\% \\
NK & 0.619 & 0.581 & 0.548 & 0.575 & 0.448 & 0.399 & \textbf{0.686} & +10.9\% \\
\bottomrule
\end{tabular}
\end{table}

\begin{table}[htbp]
\centering
\caption{Spearman rank correlation (\(\rho\)) between predicted and actual gene expression profiles for glioblastoma patients. Higher values reflect stronger consistency in gene expression ordering. Rightmost column reports the improvement of CFM-GP over CoupleVAE.}
\label{tab:glioblastoma_spearman_results}
\begin{tabular}{lccccccc|c}
\toprule
\textbf{Cell Type} & \textbf{CoupleVAE} & \textbf{scPreGAN} & \textbf{trVAE} & \textbf{scDist} & \textbf{scGen} & \textbf{CVAE} & \textbf{CFM-GP} & \textbf{Improvement (\%)} \\
\midrule
PW029 & 0.896 & 0.804 & 0.899 & 0.872 & 0.932 & 0.717 & \textbf{0.963} & +7.5\% \\
PW030 & 0.882 & 0.824 & 0.863 & 0.787 & 0.763 & 0.903 & \textbf{0.966} & +9.5\% \\
PW032 & 0.903 & 0.798 & 0.743 & 0.895 & 0.810 & 0.646 & \textbf{0.917} & +1.5\% \\
PW034 & 0.932 & 0.875 & 0.920 & 0.918 & 0.864 & 0.743 & \textbf{0.979} & +5.0\% \\
PW036 & 0.910 & 0.958 & 0.957 & 0.943 & 0.915 & 0.756 & \textbf{0.976} & +7.3\% \\
PW052 & 0.861 & 0.864 & 0.864 & 0.883 & 0.786 & 0.686 & \textbf{0.936} & +8.7\% \\
\bottomrule
\end{tabular}
\end{table}

\begin{table}[htbp]
\centering
\caption{Spearman rank correlation (\(\rho\)) between predicted and actual gene expression profiles for lupus dataset. Higher values indicate better preservation of gene expression rankings. Rightmost column reports the improvement of CFM-GP over CoupleVAE.}
\label{tab:lupus_spearman_results}
\begin{tabular}{lccccccc|c}
\toprule
\textbf{Cell Type} & \textbf{CoupleVAE} & \textbf{scPreGAN} & \textbf{trVAE} & \textbf{scDist} & \textbf{scGen} & \textbf{CVAE} & \textbf{CFM-GP} & \textbf{Improvement (\%)} \\
\midrule
B cells & 0.608 & 0.597 & 0.547 & 0.474 & 0.546 & 0.542 & \textbf{0.652} & +7.2\% \\
CD14+ Monocytes & 0.699 & 0.693 & 0.594 & 0.535 & 0.476 & 0.366 & \textbf{0.714} & +2.1\% \\
CD4 T cells & 0.496 & 0.532 & 0.533 & 0.477 & 0.417 & 0.258 & \textbf{0.562} & +13.3\% \\
CD8 T cells & 0.593 & 0.558 & 0.560 & 0.606 & 0.496 & 0.487 & \textbf{0.633} & +6.7\% \\
Dendritic cells & 0.534 & 0.525 & 0.450 & 0.524 & 0.429 & 0.271 & \textbf{0.580} & +8.6\% \\
FCGR3A+ Monocytes & 0.615 & 0.608 & 0.531 & 0.550 & 0.510 & 0.593 & \textbf{0.633} & +2.9\% \\
Megakaryocytes & 0.478 & 0.540 & 0.431 & 0.332 & 0.523 & 0.420 & \textbf{0.557} & +16.5\% \\
NK cells & 0.627 & 0.570 & 0.577 & 0.523 & 0.369 & 0.598 & \textbf{0.657} & +4.8\% \\
\bottomrule
\end{tabular}
\end{table}

\begin{table}[htbp]
\centering
\caption{Spearman rank correlation (\(\rho\)) between predicted and actual gene expression profiles for Statefate progenitor populations. Higher values indicate better preservation of gene expression rankings. Rightmost column reports the improvement of CFM-GP over CoupleVAE.}
\label{tab:statefate_spearman_results}
\begin{tabular}{lccccccc|c}
\toprule
\textbf{Cell Type} & \textbf{CoupleVAE} & \textbf{scPreGAN} & \textbf{trVAE} & \textbf{scDist} & \textbf{scGen} & \textbf{CVAE} & \textbf{CFM-GP} & \textbf{Improvement (\%)} \\
\midrule
Lin-Kit+Sca1+ & 0.783 & 0.697 & 0.699 & 0.705 & 0.707 & 0.659 & \textbf{0.805} & +2.8\% \\
Lin-Kit+Sca1- & 0.849 & 0.856 & 0.835 & 0.854 & 0.819 & 0.692 & \textbf{0.898} & +5.8\% \\
\bottomrule
\end{tabular}
\end{table}

\subsection*{Pathway Analysis Reveals Strong Biological Consistency with Real Data Through Normalized Enrichment Scores (NES)}

Pathway enrichment analysis is commonly used to interpret differential gene expression results in terms of higher-level biological processes. Rather than analyzing individual gene changes, enrichment methods assess whether specific pathways—defined by curated gene sets—are statistically overrepresented among the most perturbed genes. This helps link transcriptional variation to known signaling cascades or functional modules. In our study, we applied Gene Set Enrichment Analysis (GSEA) using ranked gene lists derived from the predicted and real expression profiles for each cell type. The gene sets were obtained from MSigDB collections, including Hallmark and KEGG pathways. Enrichment was computed using the preranked GSEA method, and Normalized Enrichment Scores (NES) were used to compare predicted versus real pathway activations. This approach enabled a functional-level evaluation of whether CFM-GP accurately recovered the biological programs perturbed in each condition.

To assess whether CFM-GP’s predictions retain functional biological relevance beyond expression-level fidelity, we conducted a pathway enrichment analysis using the top differentially expressed genes (DEGs) derived from both real and predicted expression profiles across four datasets (COVID-19, PBMC, Lupus, and Statefate). Gene Set Enrichment Analysis (GSEA) was performed independently on each condition using curated hallmark and KEGG pathways, and the resulting Normalized Enrichment Scores (NES) were used for comparative evaluation.

\begin{figure}[htbp]
    \centering
    \includegraphics[width=\textwidth]{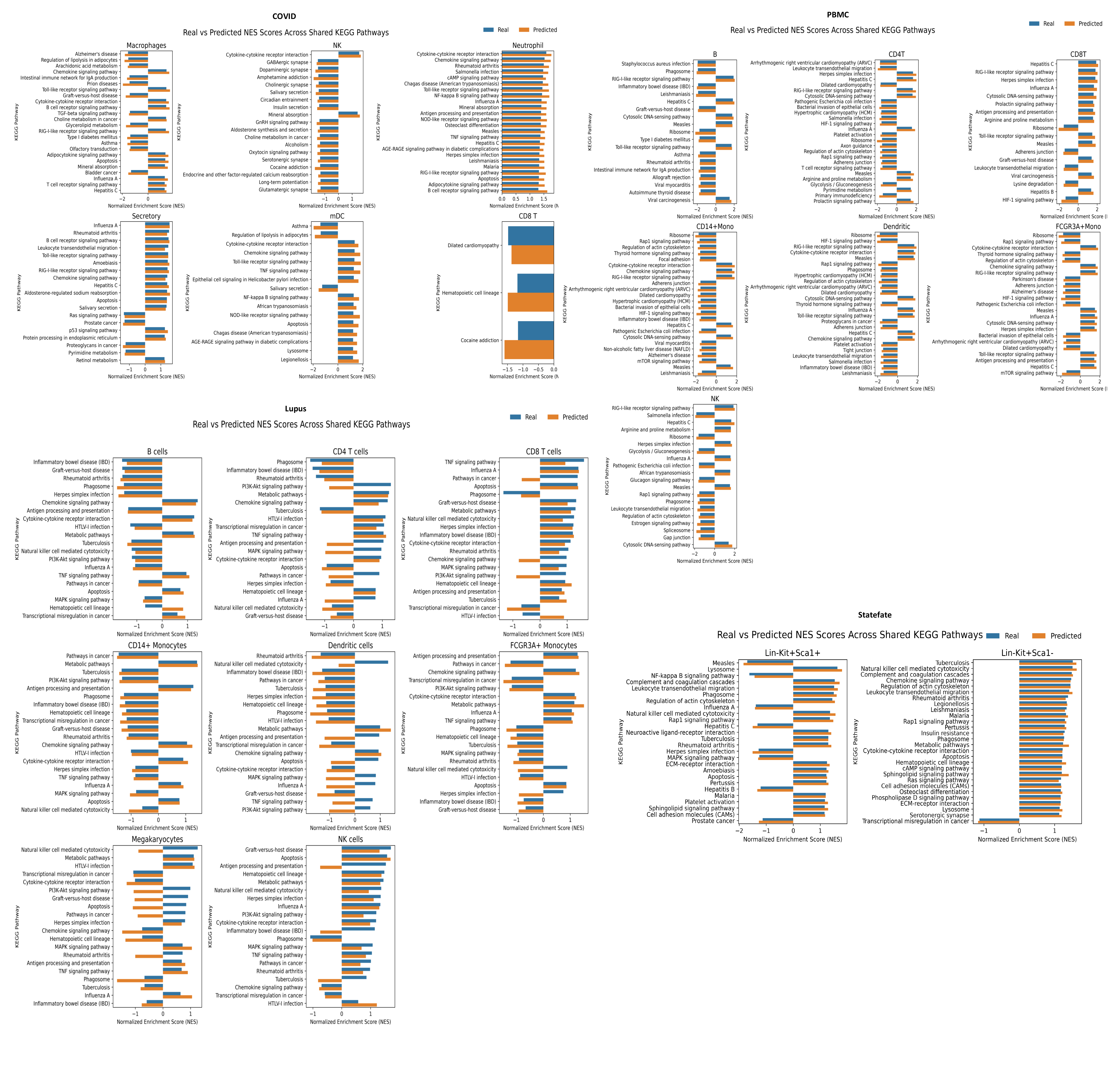}
    \caption{
Pathway enrichment analysis across four datasets (COVID-19, PBMC, Lupus, Statefate) comparing real and CFM-GP–predicted gene expression responses using normalized enrichment scores (NES). For each dataset, top pathways enriched in predicted data are compared to those in real perturbed cells, and the degree of overlap reflects the biological consistency of the model outputs.}
\label{fig:pathway_enrichment}
\end{figure}

Figure~\ref{fig:pathway_enrichment} summarizes the NES values for both real and predicted gene expression profiles across cell types and datasets. Visually, CFM-GP demonstrates strong concordance with the NES signatures of the real data, recapitulating key biological signals in both direction and magnitude. In particular, pathways associated with interferon signaling, inflammatory response, and metabolic reprogramming were consistently captured across immune-relevant cell types in COVID-19 and PBMC datasets. Similarly, in the Lupus dataset, IFN-stimulated transcriptional programs were accurately recovered, further supporting the model’s robustness to disease-specific perturbation contexts. For Statefate progenitor compartments, CFM-GP correctly preserved lineage-specific programs linked to stem and progenitor regulation, despite the relatively subtle transcriptional gradients inherent in this developmental dataset.

In the COVID-19 dataset, CFM-GP closely matched real NES profiles in macrophages, neutrophils, and mDCs, accurately recovering pathways like \textit{Toll-like receptor signaling}, \textit{cytokine interaction}, and \textit{chemokine signaling} with preserved direction and relative magnitude. Minor deviations were observed in predicted NES values for metabolic and apoptotic pathways, though overall trends aligned. In contrast, CD8 T cells showed few enriched pathways and limited recovery, reflecting weak transcriptional signal in this subset. In the PBMC dataset, CFM-GP closely tracked real NES scores across major immune subsets, particularly for \textit{interferon signaling}, \textit{cytokine interaction}, and \textit{antigen presentation} pathways in monocytes, dendritic cells, and NK cells. Directionality and magnitude were well preserved, especially in innate compartments. In CD4 and CD8 T cells, pathway concordance was more variable, with some over- or underestimation in metabolic and viral response pathways, suggesting reduced accuracy in modeling adaptive cell-specific programs. In the Lupus dataset, CFM-GP reliably recapitulated key interferon-driven transcriptional programs across B cells, monocytes, and dendritic subsets---hallmarks of systemic autoimmune activation. Predicted NES values closely mirrored real scores for pathways such as \textit{cytokine signaling}, \textit{antigen presentation}, and \textit{viral response}, with well-matched directionality and moderate magnitude deviations. Minor discrepancies emerged in more heterogeneous adaptive compartments like CD4 and CD8 T cells, where pathway-level variation was less consistently captured. In the Statefate dataset, CFM-GP preserved pathway-level regulation with notable consistency across both Lin$^-$Kit$^+$Sca1$^+$ and Lin$^-$Kit$^+$Sca1$^-$ compartments, particularly for immune-related and adhesion pathways such as \textit{leukocyte migration}, \textit{phagosome}, and \textit{CAM signaling}. While real NES values showed more polarized activation in certain contexts (e.g., \textit{MAPK} and \textit{NF-}$\kappa$\textit{B} in Sca1$^+$), the predicted scores captured directionality but tended to compress the dynamic range—suggesting the model preserved qualitative trends while smoothing quantitative extremes.

\begin{figure}[htbp]
    \centering
    \includegraphics[width=0.92\textwidth]{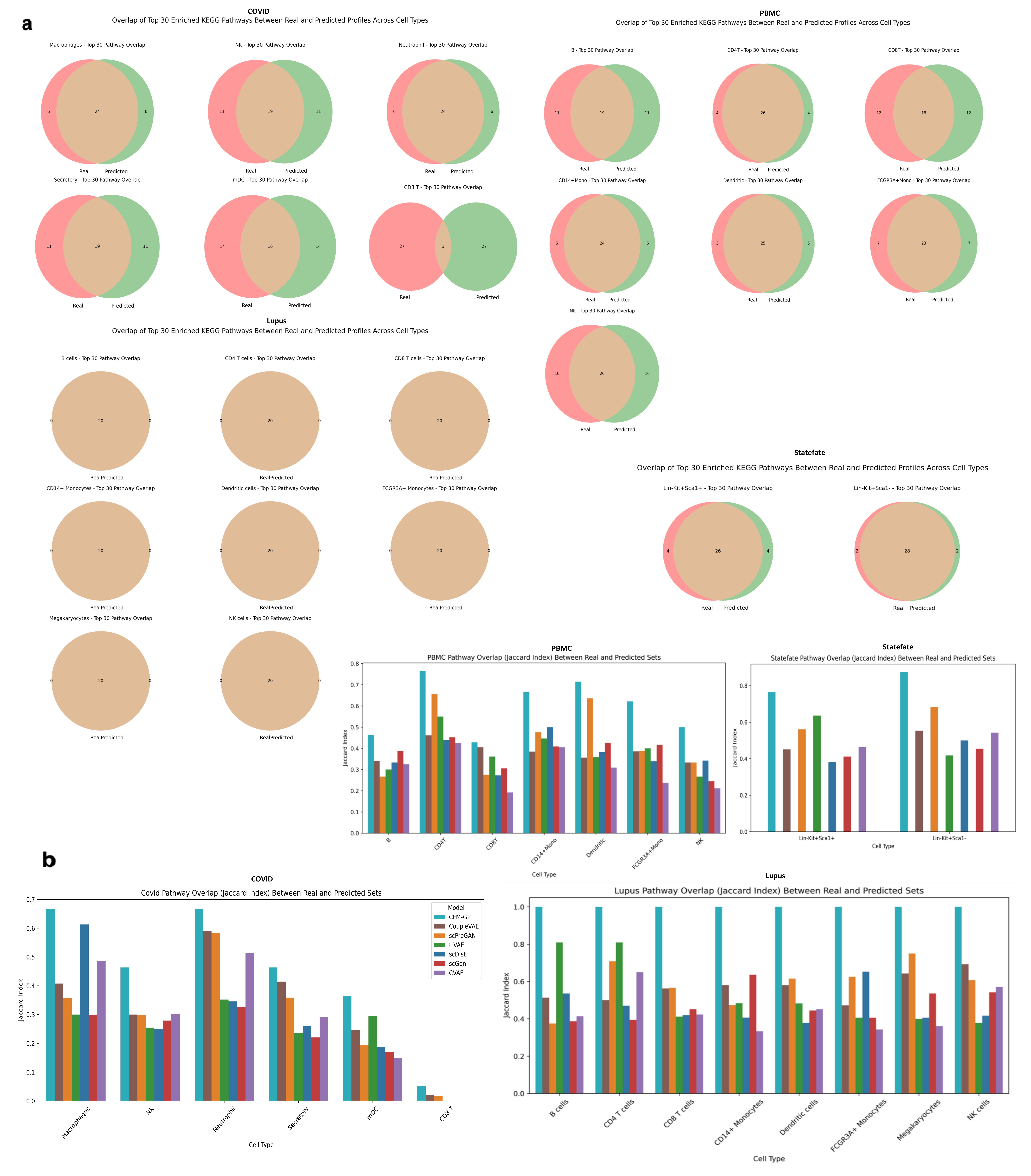}
    \caption{Pathway overlap analysis between real and predicted gene expression responses across four benchmark datasets.
(a) Venn diagrams showing the overlap between top enriched pathways (based on normalized enrichment scores, NES) derived from real and CFM-GP–predicted gene expression profiles. Pathway recovery was complete in lupus, substantial in IFN-$\beta$–stimulated PBMC and COVID-19, and moderate in cytokine-treated progenitors (Statefate), reflecting dataset complexity and developmental variation.
(b) Bar plots display Jaccard similarity scores across cell types and methods, quantifying how well each model preserved pathway-level responses. CFM-GP consistently shows the highest alignment with ground-truth signatures, particularly in innate immune subsets, indicating improved biological fidelity over competing generative models.}

\label{fig:venn_pathway_overlap}
\end{figure}

To quantify the overlap between predicted and real pathway hits, we constructed Venn diagrams comparing the top 10 enriched pathways per cell type (Figure~\ref{fig:venn_pathway_overlap}a). CFM-GP recovered all biologically enriched pathways in the Lupus dataset, showing perfect overlap with real signals. COVID-19 and PBMC datasets also exhibited substantial intersection, with minor divergence arising from predicted pathways that were plausible but not top-ranked in the real data. In the Statefate dataset, a moderate degree of overlap was observed—likely a reflection of the lower signal-to-noise ratio and fewer pathway-level perturbations typical of progenitor cells.

We used Jaccard similarity to quantify the overlap between predicted and real top-ranked pathways per cell type, computed as the ratio of shared pathways to the union of predicted and ground-truth sets (Figure~\ref{fig:venn_pathway_overlap}b). This allowed us to evaluate how well each model preserved biologically meaningful pathway signatures.

In the COVID data set, CFM-GP maintains a strong alignment of the pathway in key innate immune cells such as Macrophages and Neutrophils (Jaccard = 0.67 for both), indicating the robustness to capture infection-induced transcriptional programs. In contrast, all models struggle with adaptive compartments, with CD8 T cells showing near-zero overlap across the board, reflecting limitations in modeling T cell–specific responses under severe perturbation. In the PBMC dataset, CFM-GP consistently outperforms other models across all cell types, achieving the highest Jaccard index in most cases, particularly for CD4T and Dendritic cells. This suggests that its feature-aware design may better capture biological pathway coherence. In contrast, models like CVAE and scGen underperform on immune subsets, indicating limitations in preserving functional gene sets under domain shift. In the Lupus dataset, CFM-GP consistently achieves perfect pathway recovery (Jaccard = 1.0) across all immune cell types, suggesting exceptional robustness in capturing cell-type–specific gene sets under disease perturbation. Meanwhile, other models show variable performance, particularly weaker overlap in monocyte and dendritic subsets, highlighting challenges in modeling pathway integrity for myeloid populations under domain shift. In the Statefate dataset, CFM-GP demonstrates the strongest pathway fidelity, achieving Jaccard scores of 0.76 and 0.88 for Lin$^-$Kit$^+$Sca1$^+$ and Lin$^-$Kit$^+$Sca1$^-$ populations, respectively. Notably, while scPreGAN and trVAE perform moderately well for Lin$^-$Kit$^+$Sca1$^+$ (0.56 and 0.64), their accuracy drops for Lin$^-$Kit$^+$Sca1$^-$, highlighting potential challenges in generalizing across early hematopoietic states.

\subsection*{CFM-GP Exhibits Strong Out-of-Distribution Performance on Cross-Species Data}

Robust generalization across species is essential for modeling evolutionarily conserved gene perturbation programs and enabling translational inference. To evaluate whether CFM-GP can capture transferable response patterns, we assessed its performance in a cross-species perturbation setting using data from primary mononuclear phagocytes collected across four mammalian species: mouse, pig, rabbit, and rat. All cells were stimulated with lipopolysaccharide (LPS) for 6 hours, generating paired gene expression profiles for both control and perturbed states. This setup mirrors biological conditions where different species respond to the same stimulus with both conserved and divergent transcriptional signatures.

We define the cross-species perturbation task as an out-of-distribution (OOD) generalization problem. Specifically, models are trained on all cells from one source species (control + perturbed) and evaluated on held-out target species. This formulation tests whether perturbation dynamics learned in one species can accurately predict gene expression changes in another, given only control data from the target. CFM-GP’s performance was benchmarked against CoupleVAE across all 12 possible source-target combinations. Table~\ref{tab:cross_species_r2} reports $R^2$ values for both models alongside their relative improvement.

CFM-GP achieved higher predictive accuracy in 9 of the 12 cross-species transfers. The largest gains were observed in transfers from rat$\rightarrow$mouse (+8.1\%), rabbit$\rightarrow$mouse (+4.9\%), and mouse$\rightarrow$rabbit (+4.5\%). These results suggest that CFM-GP captures generalizable perturbation directions that extend across species boundaries. However, in three settings—most notably pig$\rightarrow$rabbit (–6.4\%)—CoupleVAE marginally outperformed CFM-GP, likely due to subtle alignment mismatches or limited divergence in perturbation signatures where latent alignment suffices.

\begin{figure}[htbp]
    \centering
    \includegraphics[width=0.9\textwidth]{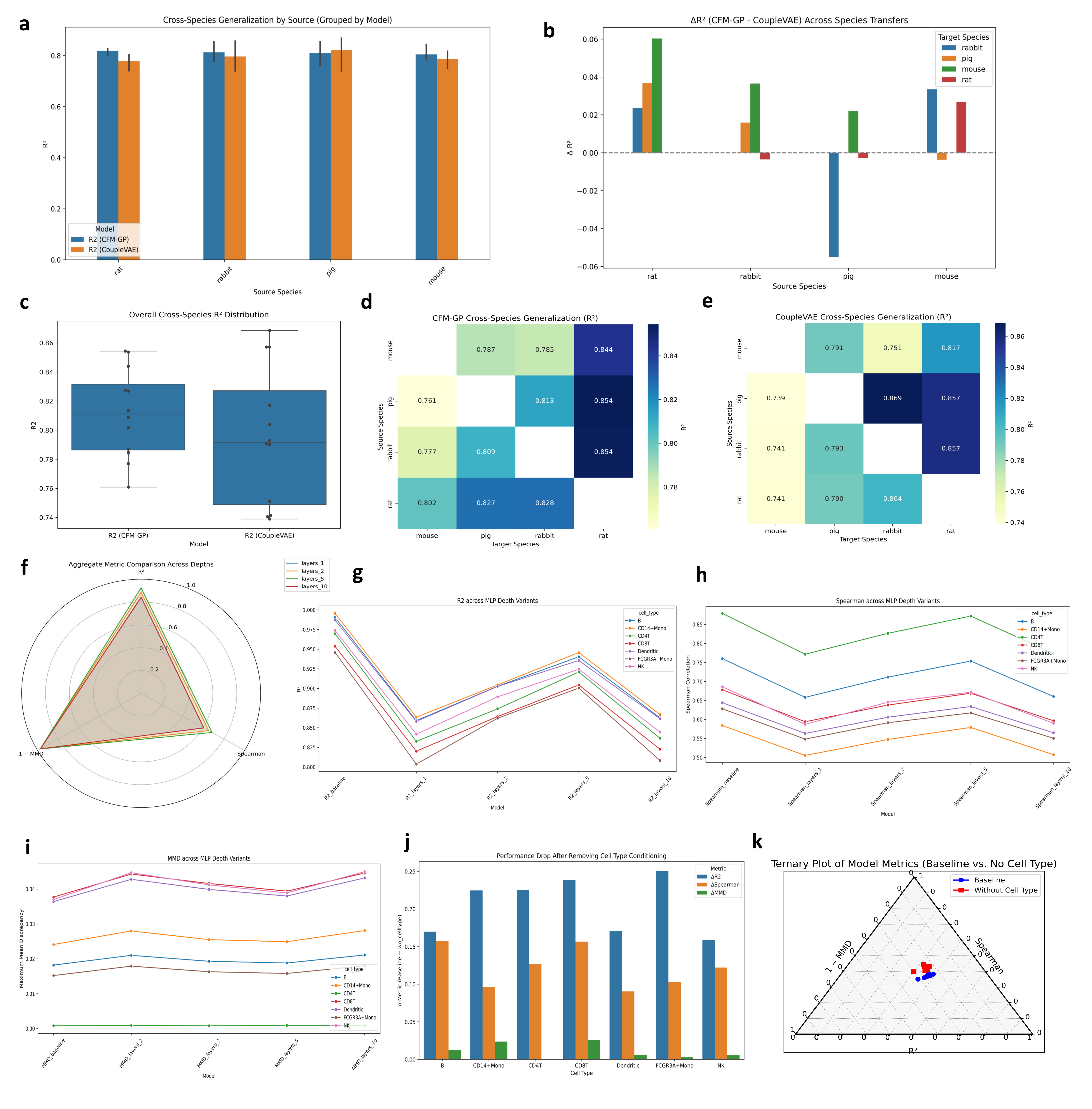}
    \caption{Cross-species generalization performance comparison between CFM-GP and CoupleVAE.
    (a) Grouped barplot showing mean $R^2$ per source species across all targets. 
    (b) $\Delta R^2$ barplot comparing CFM-GP and CoupleVAE across individual source-target species pairs. 
    (c) Boxplot showing the distribution of $R^2$ values aggregated over all transfers. 
    (d, e) Heatmaps showing per-pair $R^2$ values for CFM-GP and CoupleVAE respectively, where rows represent source species and columns represent target species.
    (f) Aggregate performance of CFM-GP across MLP depths (baseline excluded), with the 5-layer model best among ablated versions, highlighting medium-depth architectures (3–5 layers) for balanced accuracy and generalization.
    (g) R$^2$ scores per cell type across MLP depths, with baseline model retaining the highest predictive accuracy.
    (h) Spearman correlation per cell type across depths, highlighting superior rank preservation by medium-depth architectures.
    (i) Line plot to show MMD across MLP depths
    (j) Performance drop across cell types without cell type conditioning, with largest declines in R$^2$ and Spearman correlation.
    (k) Ternary plot showing metric trade-offs with and without cell type conditioning, revealing systematic shifts toward $1{-}\mathrm{MMD}$ and away from R$^2$ and Spearman, consistent with panel (j).}
    \label{fig:cross_species_combined}
\end{figure}

Panel~(a) of Figure~\ref{fig:cross_species_combined} shows model-averaged $R^2$ scores grouped by source species. CFM-GP consistently maintains higher or comparable means for each species, indicating robustness to domain shift regardless of origin. Interestingly, while CoupleVAE trails by a few percentage points in most settings, the margin is larger when the source is rat or mouse—species with more consistent perturbation signals—possibly due to CFM-GP’s ability to model continuous transformation flows rather than rely on latent sampling alone. Panel~(b) presents a detailed $\Delta R^2$ comparison between the two models. Each bar shows the relative improvement (or decline) of CFM-GP over CoupleVAE for a specific source-target pair. Most bars lie above zero, reflecting CFM-GP’s stronger transfer performance. Notably, the largest improvements occur when the source species is phylogenetically more distant from the target (e.g., rat$\rightarrow$mouse), where modeling continuous conditional dynamics may provide an advantage over static latent interpolation. Panel~(c) summarizes the overall $R^2$ distribution across all 12 transfers. The median for CFM-GP is visibly higher than CoupleVAE, and the spread is tighter. This suggests that not only does CFM-GP achieve stronger mean performance, but it is also more consistent across diverse conditions—a desirable trait for real-world generalization. Panel~(d) breaks down individual source-to-target transfer accuracy into two heatmaps, one for each model. While both methods perform well on the diagonal (source and target species being identical), off-diagonal entries reveal clearer differences. For example, CFM-GP maintains high $R^2$ when predicting from mouse to pig (0.7869) or rat to rabbit (0.8275), where CoupleVAE lags behind. In contrast, CoupleVAE shows slightly better performance from pig to rabbit and rat—cases where the biological variance is relatively lower and may not require modeling dynamic transformation.

Importantly, these results indicate that CFM-GP not only learns perturbation directions aligned with species-specific gene expression but also preserves cross-species structure in the conditional flow space. While the advantages are modest in some directions, the general trend favors CFM-GP in more divergent or biologically complex settings. This capacity is particularly valuable for comparative genomics and preclinical translational studies, where human-like predictions must be extrapolated from non-human training domains.

\begin{table}[htbp]
\centering
\caption{Cross-species generalization performance measured by \(R^2\) for CFM-GP and CoupleVAE. Higher values indicate better predictive accuracy. The rightmost column reports the percentage improvement of CFM-GP over CoupleVAE for each source-target pair.}
\label{tab:cross_species_r2}
\begin{tabular}{l l c c c}
\toprule
\textbf{Source Species} & \textbf{Target Species} & \textbf{CFM-GP} & \textbf{CoupleVAE} & \textbf{Improvement (\%)} \\
\midrule
rat    & rabbit & 0.8275 & 0.8039 & +2.94\% \\
rat    & pig    & 0.8268 & 0.7901 & +4.64\% \\
rat    & mouse  & 0.8017 & 0.7414 & +8.13\% \\
rabbit & rat    & 0.8536 & 0.8571 & \textcolor{gray}{--0.41\%} \\
rabbit & pig    & 0.8087 & 0.7928 & +2.00\% \\
rabbit & mouse  & 0.7770 & 0.7405 & +4.93\% \\
pig    & rat    & 0.8543 & 0.8571 & \textcolor{gray}{--0.33\%} \\
pig    & rabbit & 0.8134 & 0.8685 & \textcolor{gray}{--6.35\%} \\
pig    & mouse  & 0.7609 & 0.7389 & +2.97\% \\
mouse  & rat    & 0.8439 & 0.8171 & +3.29\% \\
mouse  & rabbit & 0.7846 & 0.7511 & +4.46\% \\
mouse  & pig    & 0.7869 & 0.7906 & \textcolor{gray}{--0.47\%} \\
\bottomrule
\end{tabular}
\end{table}

\subsection*{CFM-GP Shows Robust Performance Across Varying Experimental Conditions and Perturbations}

\subsection*{Ablation Study}

\subsubsection*{Impact of MLP Depth on Model Performance}
\begin{table}[htbp]
\centering
\caption{Effect of varying MLP depth in CFM-GP. We report performance across seven PBMC cell types using $R^2$, MMD, and Spearman correlation. The baseline model uses the default MLP configuration (3 layers). Comparisons are made with 1, 2, 5, and 10-layer MLP variants.}
\label{tab:mlp_depth_ablation}
\resizebox{\textwidth}{!}{
\begin{tabular}{lccc|ccc|ccc|ccc|ccc}
\toprule
\multirow{2}{*}{\textbf{Cell Type}} 
& \multicolumn{3}{c|}{\textbf{Baseline}} 
& \multicolumn{3}{c|}{\textbf{1 Layer}} 
& \multicolumn{3}{c|}{\textbf{2 Layers}} 
& \multicolumn{3}{c|}{\textbf{5 Layers}} 
& \multicolumn{3}{c}{\textbf{10 Layers}} \\
\cmidrule(lr){2-4} \cmidrule(lr){5-7} \cmidrule(lr){8-10} \cmidrule(lr){11-13} \cmidrule(lr){14-16}
& $R^2$ & MMD & Spearman 
& $R^2$ & MMD & Spearman 
& $R^2$ & MMD & Spearman 
& $R^2$ & MMD & Spearman 
& $R^2$ & MMD & Spearman \\
\midrule
B & 0.9903 & 0.0182 & 0.760 & 0.8590 & 0.0210 & 0.6582 & 0.9028 & 0.0193 & 0.7113 & 0.9403 & 0.0188 & 0.7536 & 0.8626 & 0.0211 & 0.6605 \\
CD14+Mono & 0.9955 & 0.0241 & 0.584 & 0.8635 & 0.0280 & 0.5052 & 0.9045 & 0.0255 & 0.5473 & 0.9457 & 0.0249 & 0.5790 & 0.8668 & 0.0281 & 0.5072 \\
CD4T & 0.9696 & 0.0008 & 0.879 & 0.8327 & 0.0009 & 0.7716 & 0.8740 & 0.0008 & 0.8267 & 0.9210 & 0.0009 & 0.8720 & 0.8366 & 0.0009 & 0.7738 \\
CD8T & 0.9537 & 0.0377 & 0.678 & 0.8201 & 0.0443 & 0.5940 & 0.8642 & 0.0416 & 0.6378 & 0.9045 & 0.0394 & 0.6690 & 0.8226 & 0.0446 & 0.5965 \\
Dendritic & 0.9870 & 0.0364 & 0.644 & 0.8581 & 0.0428 & 0.5631 & 0.9026 & 0.0399 & 0.6059 & 0.9356 & 0.0380 & 0.6340 & 0.8615 & 0.0432 & 0.5649 \\
FCGR3A+Mono & 0.9457 & 0.0152 & 0.628 & 0.8037 & 0.0179 & 0.5482 & 0.8621 & 0.0163 & 0.5914 & 0.9005 & 0.0158 & 0.6172 & 0.8083 & 0.0180 & 0.5505 \\
NK & 0.9739 & 0.0370 & 0.686 & 0.8418 & 0.0447 & 0.5873 & 0.8893 & 0.0412 & 0.6454 & 0.9245 & 0.0389 & 0.6717 & 0.8444 & 0.0450 & 0.5899 \\
\bottomrule
\end{tabular}
}
\end{table}

To evaluate the sensitivity of CFM-GP to architectural complexity, we systematically varied the depth of the shared MLP from 1 to 10 layers while keeping all other components fixed. This ablation isolates the effect of nonlinear capacity on prediction accuracy, distributional alignment (MMD), and rank preservation (Spearman) (Table ~\ref{tab:mlp_depth_ablation}). Our goal was to identify whether deeper networks improve modeling of complex gene perturbation patterns, or if excessive depth introduces overfitting or degradation.

Figure~\ref{fig:cross_species_combined}f presents the aggregate performance of CFM-GP across varying MLP depths, using a radar plot of $R^2$, Spearman correlation, and $1 - \mathrm{MMD}$. The baseline 3-layer configuration consistently achieves the highest scores across all three metrics, demonstrating an effective balance between accuracy and generalization. The 5-layer variant ranks second overall and performs best among the ablated configurations. Shallower networks (e.g., 1-layer) perform notably worse, while very deep networks (e.g., 10-layer) offer limited gains, likely due to overfitting. These results highlight medium-depth architectures, particularly the 3- and 5-layer models, as the most effective design choices.

Figure~\ref{fig:cross_species_combined}g illustrates the R² scores across varying MLP depths for each cell type. The baseline consistently achieves the highest R², as expected from the fully tuned model. Among ablated versions, the 5-layer model consistently recovers the most predictive accuracy across all cell types. Shallower (1-layer) and deeper (10-layer) configurations notably underperform, with the 1-layer variant exhibiting the steepest drop, especially in FCGR3A+ Mono and CD8T cells. These trends support baseline configuration as the most balanced depth for retaining predictive fidelity across cellular contexts.

Figure~\ref{fig:cross_species_combined}h tracks Spearman correlation across MLP depth variants for each cell type, providing insight into the model's ability to preserve gene expression ordering. The baseline model achieves the highest rank fidelity overall, with the 5-layer variant again coming closest to recovering baseline performance. A single hidden layer substantially degrades correlation across all cell types—most notably in CD4T and B cells—while moderate recovery is observed as depth increases to 2 and 5 layers. The 10-layer variant shows declining trends again, particularly in low-signal populations like CD14+Mono and FCGR3A+Mono, suggesting overparameterization may destabilize rank structure preservation.

Figure~\ref{fig:cross_species_combined}i presents MMD across varying MLP depths, offering a distributional perspective on how well predicted and real gene profiles align. The baseline model consistently yields the lowest MMD values, confirming its strong distributional calibration. Shallow variants (e.g., 1-layer) lead to notable increases in MMD for most cell types—especially CD14+Mono, CD8T, and NK—indicating weaker distribution matching. Deeper configurations (layers 5 and 10) partially recover performance, but never match the original baseline. Interestingly, CD4T maintains near-zero MMD across all depths, suggesting either inherently easy-to-align distributions or metric saturation. Overall, the figure highlights that while moderate depth can stabilize performance, deeper networks risk diminishing returns or increased mismatch in more complex populations.

\subsubsection*{Importance of Cell Type Conditioning for Accurate Perturbation Modeling
}
\begin{table}[htbp]
\centering
\caption{Ablation analysis of cell type conditioning in CFM-GP. Performance is reported across seven PBMC cell types with and without cell type conditioning. Metrics include coefficient of determination ($R^2$), Maximum Mean Discrepancy (MMD), and Spearman rank correlation.}
\label{tab:ablation_celltype}
\begin{tabular}{lccc|ccc}
\toprule
\multirow{2}{*}{\textbf{Cell Type}} & \multicolumn{3}{c|}{\textbf{Baseline (w/ Cell Type)}} & \multicolumn{3}{c}{\textbf{Without Cell Type}} \\
\cmidrule(lr){2-4} \cmidrule(lr){5-7}
& $R^2$ & MMD & Spearman & $R^2$ & MMD & Spearman \\
\midrule
B & 0.9903 & 0.0182 & 0.760 & 0.8205 & 0.0310 & 0.6026 \\
CD14+Mono & 0.9955 & 0.0241 & 0.584 & 0.7708 & 0.0478 & 0.4874 \\
CD4T & 0.9696 & 0.0008 & 0.879 & 0.7441 & 0.0009 & 0.7518 \\
CD8T & 0.9537 & 0.0377 & 0.678 & 0.7153 & 0.0637 & 0.5214 \\
Dendritic & 0.9870 & 0.0364 & 0.644 & 0.8163 & 0.0425 & 0.5536 \\
FCGR3A+Mono & 0.9457 & 0.0152 & 0.628 & 0.6949 & 0.0180 & 0.5251 \\
NK & 0.9739 & 0.0370 & 0.686 & 0.8151 & 0.0424 & 0.5639 \\
\bottomrule
\end{tabular}
\end{table}

We ablated the cell type conditioning component in the cross-attention and decoder modules to assess its role in guiding gene expression prediction. By training a variant of CFM-GP without explicit access to cell identity, we tested whether performance relies on cell type context to model heterogeneous responses shown in Table ~\ref{tab:ablation_celltype}. This experiment directly probes the model’s ability to generalize across cell types and uncovers the contribution of biological conditioning to multi-metric robustness.

Figure~\ref{fig:cross_species_combined}j illustrates the performance degradation across cell types when cell type conditioning is removed from CFM-GP. The most substantial impact is observed in the R\textsuperscript{2} metric (blue bars), where predictive accuracy drops by over 0.22 for several populations including CD4T, CD8T, and FCGR3A+Mono, highlighting the model's strong reliance on cell type context to model accurate gene responses. Spearman rank correlation (orange bars) also suffers, with a maximum decline of approximately 0.16, especially for B and CD8T cells, indicating a weakened ability to preserve gene ordering. Interestingly, the $\Delta$MMD values (green bars) remain relatively low, suggesting that although expression ranking and magnitude suffer, the overall distributional alignment degrades less severely. This decoupling underscores that MMD alone may mask biologically important shifts in gene-specific behavior. 

Figure~\ref{fig:cross_species_combined}k presents a ternary plot comparing the trade-offs among three key evaluation metrics—R\textsuperscript{2}, Spearman correlation, and $1{-}\text{MMD}$—for CFM-GP models trained with (blue dots) and without (red squares) cell type conditioning. Each point represents a normalized performance profile for a specific cell type, where proximity to a vertex indicates stronger emphasis on that particular metric. The plot reveals a systematic shift in the red squares (ablated model) away from the R\textsuperscript{2} and Spearman corners and slightly toward the $1{-}\text{MMD}$ vertex. This shift implies that removing cell type conditioning results in a notable degradation in predictive accuracy and rank preservation, consistent with the bar plot in Figure~\ref{fig:cross_species_combined}j. At the same time, $1{-}\text{MMD}$ remains relatively stable, possibly because it reflects global distributional alignment, which is less sensitive to fine-grained gene-wise structure.
Interestingly, some red squares remain close to their baseline counterparts, suggesting a subset of cell types may be less dependent on explicit cell type features for general structure modeling. However, the consistent directional shift across most cell types highlights that cell type conditioning is critical for achieving a balanced optimization across all three axes—especially preserving nuanced gene relationships and ensuring biologically faithful predictions.

\section*{Conclusion}
In this work, we introduce CFM-GP, a unified and scalable framework for predicting gene expression responses to perturbations across diverse cell types and species. By leveraging conditional flow matching, CFM-GP models perturbation dynamics as time-dependent vector fields, enabling biologically grounded, continuous transformations from control to perturbed states. Our extensive evaluations across five biologically diverse datasets and a cross-species transfer setting demonstrate that CFM-GP consistently outperforms existing baselines in predictive accuracy, distributional alignment, and pathway-level recovery.
Despite its strengths, CFM-GP has several limitations. First, while it generalizes across cell types seen during training, its performance on completely novel perturbations remains unexplored and may be constrained. Second, the model assumes a smooth and continuous transition between control and perturbed states, which might not hold in cases of abrupt or bifurcated cellular responses. Third, we did not analyze training efficiency or computational cost, which is critical for practical deployment in large-scale perturbation screening. Finally, although the model implicitly captures cell-type specificity through conditioning, explicit incorporation of prior biological networks or regulatory constraints may further enhance interpretability and robustness.

Future work will explore these directions by integrating structural priors, evaluating zero-shot perturbation settings, and optimizing the architecture for greater computational efficiency and uncertainty quantification. Nonetheless, CFM-GP represents a significant step toward scalable, biologically faithful modeling of perturbation responses, with broad applications in functional genomics, drug discovery, and precision medicine.

\section*{Methods}

\subsection*{Problem Formulation}

Modeling the effects of gene perturbation at single-cell resolution is a central challenge in systems biology. Given the control (unperturbed) gene expression profile $\mathbf{x}_c \in \mathbb{R}^G$ of an individual cell and its corresponding cell type $c \in \mathcal{C}$, our objective is to predict the perturbed gene expression profile $\mathbf{y} \in \mathbb{R}^G$ following a perturbation shown in Figure ~\ref{fig:method_pipeline}. Here, $G$ denotes the number of genes, and $\mathcal{C}$ is the set of cell types under study. Formally, we aim to learn the conditional distribution: $p(\mathbf{y} \mid \mathbf{x}_c, c),$

where $p$ models the distribution of perturbed gene expression conditioned on both the control expression and cell type. 
Traditional models for gene perturbation typically learn the distribution $p(\mathbf{y} \mid \mathbf{x}_c)$ \emph{separately for each cell type}. As a result, they require training and maintaining a distinct model for every cell type $c \in \mathcal{C}$, which greatly limits scalability and cross-cell-type generalization. In contrast, our approach seeks to directly model $p(\mathbf{y} \mid \mathbf{x}_c, c)$ with a single unified model, thereby enabling joint learning across all cell types and more efficient utilization of data.

\subsection*{Conditional Flow Matching}

We propose to model the conditional transformation from control to perturbed gene expression via \emph{Conditional Flow Matching} (CFM), which learns a continuous-time vector field parameterized by a neural network~\cite{lipman2022flow}. This vector field describes the transformation of $\mathbf{x}_c$ to $\mathbf{y}$, explicitly conditioned on the cell type $c$.

Let $v_\theta(\mathbf{x}, t \mid \mathbf{x}_c, c)$ denote a neural vector field parameterized by $\theta$, which governs the flow from control to perturbed expression over a continuous time variable $t \in [0, 1]$. The transformation is defined as an ordinary differential equation (ODE):

\begin{equation}
    \frac{d\mathbf{x}(t)}{dt} = v_\theta(\mathbf{x}(t), t \mid \mathbf{x}_c, c), \qquad \mathbf{x}(0) = \mathbf{x}_c.
\end{equation}

The goal is to find $v_\theta$ such that integrating the flow from $t=0$ to $t=1$ transports the control expression $\mathbf{x}_c$ to a sample drawn from the perturbed distribution $p(\mathbf{y} \mid \mathbf{x}_c, c)$.

\subsubsection*{Training Objective}

During training, we assume access to paired observations $\{(\mathbf{x}_c^{(i)}, \mathbf{y}^{(i)}, c^{(i)})\}_{i=1}^N$, where $(\mathbf{x}_c^{(i)}, \mathbf{y}^{(i)})$ are control and perturbed gene expression profiles of the same cell type $c^{(i)}$. For each pair, we construct interpolated states:
\begin{equation}
    \boldsymbol{\mu}_t = (1-t)\mathbf{x}_c + t\mathbf{y}, \qquad t \sim \mathcal{U}[0, 1].
\end{equation}
The target vector field at $\boldsymbol{\mu}_t$ is defined as:
\begin{equation}
    \mathbf{v}_{\text{target}} = \mathbf{y} - \mathbf{x}_c.
\end{equation}

The model is trained by minimizing the mean squared error between the neural vector field and the ground truth direction, across all interpolated states:
\begin{equation}
    \mathcal{L}(\theta) = \mathbb{E}_{(\mathbf{x}_c, \mathbf{y}, c), t} \left[ \left\| v_\theta(\boldsymbol{\mu}_t, t \mid \mathbf{x}_c, c) - (\mathbf{y} - \mathbf{x}_c) \right\|^2 \right].
\end{equation}

This objective enables the model to learn a continuous, cell-type-aware transformation from control to perturbed gene expression.

\subsubsection*{Inference}

At inference time, given a new control profile $\mathbf{x}_c^*$ and cell type $c^*$, we simulate the learned vector field to generate a predicted perturbed profile $\hat{\mathbf{y}}$ by integrating the ODE:
\begin{equation}
    \frac{d\mathbf{x}(t)}{dt} = v_\theta(\mathbf{x}(t), t \mid \mathbf{x}_c^*, c^*), \qquad \mathbf{x}(0) = \mathbf{x}_c^*.
\end{equation}
The predicted perturbed profile is taken as $\hat{\mathbf{y}} = \mathbf{x}(1)$. In practice, the ODE is solved by explicit forward Euler integration: the interval $[0, 1]$ is discretized into uniform steps, and at each step, the state is updated by moving in the direction of the learned vector field.  This procedure enables fast, cell-type-specific simulation of gene expression responses to perturbations, using only control-state expression and a single, unified model.

\subsection*{Implementation Details}

In our implementation, the conditional vector field $v_\theta$ is parameterized by a fully connected neural network that takes as input the current state, control state, time, and cell type, and outputs a velocity vector in gene expression space. Specifically, the input to the network at each time $t$ consists of: the current interpolated gene expression vector $\mathbf{x}_t \in \mathbb{R}^G$, the original control gene expression vector $\mathbf{x}_c \in \mathbb{R}^G$, an embedding vector for the cell type $c$, and a time embedding for $t$. These vectors are concatenated to form a single input vector. The cell type $c$ is embedded via a learnable embedding layer that maps each categorical cell type index to a $d_c$-dimensional vector (we set $d_c=16$). The continuous time variable $t$ is projected to a $d_t$-dimensional embedding (with $d_t=16$) using a linear layer followed by a ReLU nonlinearity.

The concatenated input $[\mathbf{x}_t, \mathbf{x}_c, \text{Embed}(c), \text{Embed}(t)]$ is passed through a multi-layer perceptron (MLP) comprising three fully connected layers. The first two layers have $256$ hidden units each, each followed by a ReLU activation. The final output layer projects the hidden representation back to $\mathbb{R}^G$, matching the dimension of the gene expression space. Formally, the MLP can be described as:
\begin{align}
    h_0 &= \text{Concat}(\mathbf{x}_t, \mathbf{x}_c, \text{Embed}(c), \text{Embed}(t)) \\
    h_1 &= \text{ReLU}(W_1 h_0 + b_1) \\
    h_2 &= \text{ReLU}(W_2 h_1 + b_2) \\
    v_\theta(\mathbf{x}_t, t \mid \mathbf{x}_c, c) &= W_3 h_2 + b_3
\end{align}
where $W_i, b_i$ denote the weights and biases of each layer. Finally, the model was trained for $50$ epochs using the Adam~\cite{adam} optimizer with a learning rate of $1 \times 10^{-4}$ and  mean squared error loss as described previously. Mini-batch stochastic gradient descent was employed with a batch size of $32$ samples. All experiments were conducted five times and on a cluster equipped with four NVIDIA RTX
 A4500 GPUs.

\subsection*{Datasets}

\subsubsection*{COVID-19 Peripheral Airway Dataset}

We used a publicly available single-cell RNA-seq dataset profiling the peripheral airway epithelium and immune cells from COVID-19 patients and healthy controls \cite{lotfollahi2022mapping}, available under accession code \href{https://www.ncbi.nlm.nih.gov/geo/query/acc.cgi?acc=GSE145926}{GSE145926}. The dataset contains 4713 training, 602 validation, and 556 test cells, each with 6000 genes. Six immune and epithelial cell types are represented: CD8 T cells, macrophages, NK cells, neutrophils, secretory cells, and myeloid dendritic cells (mDCs). Each split includes paired control and perturbed profiles per cell, enabling direct modeling of transcriptional shifts. Macrophages and neutrophils are the most abundant cell types, accounting for the majority of paired samples across all data splits.

\subsubsection*{PBMC Stimulation Dataset}

We also used single-cell RNA-seq dataset of peripheral blood mononuclear cells (PBMCs) stimulated ex vivo, available under accession code \href{https://www.ncbi.nlm.nih.gov/geo/query/acc.cgi?acc=GSE96583}{GSE96583}. The dataset contains 5998 training, 765 validation, and 2924 test cells, each with 6998 genes. It covers seven major immune cell types: B cells, CD4 and CD8 T cells, CD14\textsuperscript{+} monocytes, FCGR3A\textsuperscript{+} monocytes, dendritic cells, and NK cells. Each cell has paired control and stimulated profiles, enabling precise modeling of perturbation effects. CD4 T cells are the most abundant, followed by monocytes and B cells, offering diverse signal across innate and adaptive immune compartments.

\subsubsection*{Glioblastoma Drug Response Dataset}

To evaluate perturbation prediction in a cancer context, we utilized a single-cell RNA-seq dataset profiling glioblastoma patient-derived cells under drug treatment \cite{zhao2021deconvolution}, available under accession code \href{https://www.ncbi.nlm.nih.gov/geo/query/acc.cgi?acc=GSE148842}{GSE148842}. The dataset includes 30{,}842 cells with 1000 genes, derived from six unique patient samples (PW029, PW030, PW032, PW034, PW036, and PW052), which we treat as distinct pseudo–cell types. Paired control and treated profiles are provided for each sample, with PW034 and PW036 comprising the largest subsets (15{,}288 and 6120 cells, respectively). The data were split into training (21{,}589 cells), validation (4626 cells), and test (4627 cells) sets, enabling evaluation of generalization across patients.

\subsubsection*{Systemic Lupus Erythematosus (SLE) Dataset}

We analyzed a peripheral blood single-cell RNA-seq dataset from lupus patients \cite{kang2018multiplexed}, sourced from \href{https://www.ncbi.nlm.nih.gov/geo/query/acc.cgi?acc=GSE96583}{GSE96583}, comprising 13{,}795 cells and 1000 genes. The dataset spans eight immune cell types, including CD4 and CD8 T cells, B cells, monocyte subsets, dendritic cells, NK cells, and megakaryocytes. We used a 70:15:15 split across training, validation, and test sets. CD4 T cells and CD14\textsuperscript{+} monocytes were the most abundant, offering a rich basis for modeling diverse immune perturbation responses.

\subsubsection*{Cytokine-Driven Progenitor Fate Dataset (Statefate)}

This dataset \cite{weinreb2020lineage}, obtained from \href{https://www.ncbi.nlm.nih.gov/geo/query/acc.cgi?acc=GSE140802}{GSE140802}, captures cytokine-induced transcriptional changes in hematopoietic progenitor cells. It includes 28{,}249 cells across two sorted populations: Lin-Kit\textsuperscript{+}Sca1\textsuperscript{+} and Lin-Kit\textsuperscript{+}Sca1\textsuperscript{--}. The data are split into training (19{,}774), validation (4237), and test (4238) sets, with over 20{,}000 paired control and perturbed cells in the Sca1\textsuperscript{+} group alone—making it well-suited for modeling lineage-specific perturbation effects.

\section*{Data Availability}
All datasets used in this study are publicly available. These datasets include COVID-19 \cite{lotfollahi2022mapping}, PBMC, Glioblastoma \cite{zhao2021deconvolution}, Lupus \cite{kang2018multiplexed}, Statefate \cite{weinreb2020lineage}. These are available from the Gene Expression Omnibus under accession codes 
\href{https://www.ncbi.nlm.nih.gov/geo/query/acc.cgi?acc=GSE145926}{GSE145926}, \href{https://www.ncbi.nlm.nih.gov/geo/query/acc.cgi?acc=GSE96583}{GSE96583},
\href{https://www.ncbi.nlm.nih.gov/geo/query/acc.cgi?acc=GSE148842 }{GSE148842},
\href{https://www.ncbi.nlm.nih.gov/geo/query/acc.cgi?acc=GSE96583}{GSE96583} and
\href{https://www.ncbi.nlm.nih.gov/geo/query/acc.cgi?acc=GSE140802}{GSE140802} respectively. The out-of-ditribution cross-spcceis dataset is available on UK Biostudies under accession
\href{https://www.ebi.ac.uk/arrayexpress/experiments/E-MTAB-6754/?query=tzachi+hagai}{E-MTAB-6754}.

\section*{Code Availability}
CFM-GP is written Python and uses standard Python libraries. The code is available at \href{https://github.com/abrarrahmanabir/CFM-GP.git}{https://github.com/abrarrahmanabir/CFM-GP.git}.

\section*{Hyperparameters for CFM-GP}

Table~\ref{tab:cfm_hyperparams} presents all the hyperparameters used to conduct the experiments of CFM-GP.

\begin{table}[htbp]
  \centering
  \caption{Key Hyperparameters}
  \label{tab:cfm_hyperparams}
  \resizebox{\textwidth}{!}{
  \begin{tabular}{llp{8.2cm}}
    \toprule
    \textbf{Hyperparameter} & \textbf{Value/Default} & \textbf{Description} \\
    \midrule
    \multicolumn{3}{l}{\textit{Data / Loader}} \\
    \midrule
    \texttt{train\_path}         & (user-specified) & Path to training dataset (.pt file) \\
    \texttt{val\_path}           & (user-specified) & Path to validation dataset (.pt file) \\
    \texttt{test\_path}          & (user-specified) & Path to test dataset (.pt file) \\
    \texttt{batch\_size}         & 32               & Mini-batch size for DataLoader \\
    \midrule
    \multicolumn{3}{l}{\textit{Model Architecture}} \\
    \midrule
    \texttt{input\_dim}          & (auto-inferred)  & Number of gene features per cell \\
    \texttt{num\_cell\_types}    & (auto-inferred)  & Number of distinct cell types \\
    \texttt{hidden\_dim}         & 256              & Width of hidden layers in MLP \\
    \texttt{cell\_embedding\_dim}& 16               & Embedding size for cell type \\
    \texttt{time\_embedding\_dim}& 16               & Embedding size for time $t$ \\
    \texttt{num\_mlp\_layers}    & 3                & Number of linear layers (excluding embeddings) \\
    \midrule
    \multicolumn{3}{l}{\textit{Training}} \\
    \midrule
    \texttt{epoch}               & 50 (default)     & Number of training epochs \\
    \texttt{lr}                  & $1 \times 10^{-4}$ & Adam learning rate \\
    \texttt{optimizer}           & Adam             & Optimizer used \\
    \texttt{loss\_fn}            & MSELoss          & Mean squared error loss function \\
    \texttt{device}              & cuda/cpu         & Device for training \\
    \texttt{save\_path}          & model.pt         & Path to save the trained model \\
    \midrule
    \multicolumn{3}{l}{\textit{Evaluation}} \\
    \midrule
    \texttt{n\_steps}            & 10               & Steps for iterative trajectory integration \\
    \texttt{out\_prefix}         & CFM              & Prefix for evaluation result files \\
    \texttt{save\_dir}           & ./results/       & Directory for evaluation output \\
    \texttt{kernel}              & rbf              & MMD kernel (for metric evaluation) \\
    \texttt{gamma} (MMD)         & 1.0              & RBF kernel gamma for MMD computation \\
    \texttt{top\_100\_DEGs}      & 100              & Top DEGs for cell-type specific R$^2$ evaluation \\
    \bottomrule
  \end{tabular}
  }
\end{table}

\bibliography{sample}



\section*{Acknowledgements}
This work was supported in part by Virginia Tech, the Department of Computer Science, and the U.S. National Science Foundation (NSF) under Awards \#2125798, \#2344169, and \#2319522. We gratefully acknowledge their support and resources that made this research possible.




\end{document}